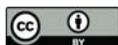

# Defining and resolving current systems in geospace

N. Y. Ganushkina[1,2], M. W. Liemohn[2], S. Dubyagin[1], I. A. Daglis[3], I. Dandouras[4], D. L. De Zeeuw[2], Y. Ebihara[5], R. Ilie[2], R. Katus[2], M. Kubyshkina[6], S. E. Milan[7,8], S. Ohtani[9], N. Ostgaard[8], J. P. Reistad[8], P. Tenfjord[8], F. Toffoletto[10], S. Zaharia[11], and O. Amariutei[1]

[1]Earth Observations Department, Finnish Meteorological Institute, Helsinki, Finland
[2]Department of Climate and Space Sciences and Engineering, University of Michigan, Ann Arbor, Michigan, USA
[3]Department of Physics, University of Athens, Panepistimiopolis Zografou, 15784 Athens, Greece
[4]Astrophysics and Planetary Science Research Institute, CNRS/University of Toulouse, Toulouse, France
[5]Research Institute for Sustainable Humanosphere, Kyoto University, Gokasho, Uji, Kyoto, Japan
[6]Institute of Physics, University of St. Petersburg, St. Petersburg, Russia
[7]Department of Physics and Astronomy, University of Leicester, Leicester, UK
[8]Birkeland Centre for Space Science, Department of Physics and Technology, University of Bergen, Bergen, Norway
[9]Johns Hopkins University Applied Physics Laboratory, Laurel, Maryland, USA
[10]Physics and Astronomy Department, Rice University, Houston, Texas, USA
[11]ISR-1 Division, Los Alamos National Laboratory, Los Alamos, New Mexico, USA

*Correspondence to:* N. Y. Ganushkina (natalia.ganushkina@fmi.fi)

**Abstract.** Electric currents flowing through near-Earth space ($R \leq 12\,R_E$) can support a highly distorted magnetic field topology, changing particle drift paths and therefore having a nonlinear feedback on the currents themselves. A number of current systems exist in the magnetosphere, most commonly defined as (1) the dayside magnetopause Chapman–Ferraro currents, (2) the Birkeland field-aligned currents with high-latitude "region 1" and lower-latitude "region 2" currents connected to the partial ring current, (3) the magnetotail currents, and (4) the symmetric ring current. In the near-Earth nightside region, however, several of these current systems flow in close proximity to each other. Moreover, the existence of other temporal current systems, such as the substorm current wedge or "banana" current, has been reported. It is very difficult to identify a local measurement as belonging to a specific system. Such identification is important, however, because how the current closes and how these loops change in space and time governs the magnetic topology of the magnetosphere and therefore controls the physical processes of geospace. Furthermore, many methods exist for identifying the regions of near-Earth space carrying each type of current. This study presents a robust collection of these definitions of current systems in geospace, particularly in the near-Earth nightside magnetosphere, as viewed from a variety of observational and computational analysis techniques. The influence of definitional choice on the resulting interpretation of physical processes governing geospace dynamics is presented and discussed.

## 1 Introduction

Electric currents in geospace can support a highly distorted magnetic field topology, changing particle drift paths and therefore having a nonlinear feedback on the currents themselves. A number of current systems exist in the magnetosphere, most notably the dayside magnetopause Chapman–Ferraro currents, high-latitude "region 1" field-aligned Birkeland currents, "region 2" field-aligned currents connected to the partial ring current, magnetotail currents, and the symmetric ring current. In addition, there are several current systems that only exist at certain times and places, further complicating the identification and understanding of the flow of electric current through geospace and its nonlinear effects on the system. In the near-Earth nightside, for instance, several of these current systems flow in close prox-imity to each other and it is very difficult to identify a local measurement as belonging to a specific system.

Such identification is important, however, because how the current closes and how these loops change in space and time governs the magnetic topology of the magnetosphere and therefore controls the physical processes of geospace. As an example, consider the feedback on the electric fields within the magnetosphere–ionosphere system. Of the various currents flowing on or near the magnetopause, the region 1 current system connects to the ionosphere and influences the convection pattern. Similarly, in the inner magnetosphere, only the partial ring current flows through the ionosphere and therefore modifies the electric potential in this region. The other current systems, however intense they might be, exert no control over the electric potential pattern in geospace.

Part of the issue is that currents are difficult to quantify in both data and modeling. While electric current is a physically observable quantity, it is difficult to obtain from measurements. Several different techniques have been created for extracting electric current from data, but all are indirect methods that involve assumptions about the state of the system. While electric current can be directly calculated from numerical model output, there are concerns about the validity of the results because of the inherent assumptions built into the modeling technique at the equation set definition, numerical scheme implementation, or output extraction and processing.

Another factor contributing to the problem is that the space physics community does not necessarily agree on the definition of the various current systems. Different studies assume particular features in the data or model results correspond to certain current systems, and the different definitions between studies for supposedly the same current lead to confusion and unnecessary controversy. Therefore, it is useful to compile a comprehensive list of the methods used to define currents and the regions in which these currents flow.

Discussing current systems puts this review within the realm of the **E**, **j** paradigm, in which electric fields and currents take the dominant role. This is in contrast to the **B**, **v** paradigm, in which magnetic field and plasma velocity are the primary quantities. While Parker (1996) and Vasyliunas (2005) made the case that the **B**, **v** paradigm is the preferred system for space plasma physics, each paradigm has its advantages and weaknesses. While the **B**, **v** paradigm is the more natural equation set for calculating bulk motion and magnetic topology in space plasmas, it can sometimes be cumbersome when trying to interpret physical processes. The **E**, **j** paradigm is often more natural in terms of gaining physical understanding, but it includes an assumption of "stationarity" in the solution. Maxwell's equations do not imply causality but simply state a relationship, and both can be useful for advancing knowledge of space physics in general and the geospace system in particular.

## 2 Analysis methods of current systems

### 2.1 Currents in space in brief

Electric currents are produced by charges in motion. Large-scale currents in space are produced by charged particles of the solar wind, magnetospheres, and ionospheres. These currents are sources of magnetic field in the regions of space. The most common and straightforward way to discuss currents in the planetary magnetospheres is to classify them as (1) boundary currents, (2) ring currents, (3) ionospheric currents, (4) field-aligned currents, and (5) magnetotail currents.

The Earth's magnetosphere boundary current, called the Chapman–Ferraro current, is produced by the solar protons and electrons which penetrate the geomagnetic field. Currents that are produced by the balance between magnetic and plasma pressure, such as at plasma boundaries and localized peaks, are diamagnetic. A ring current is due to the motion of trapped particles in an inhomogeneous magnetic field as the particles undergo gradient and curvature drifts.

The density of particles in ionospheres is high enough that collisions cannot be ignored. Collisions give rise to momentum transfer and the electrical conductivity is important. Current flows in ionospheres can be described by the generalized Ohm's law. The ionospheric plasma is also anisotropic if the planetary magnetic field is strong. If a tensor conductivity in the Ohm's law is used, it leads to the Hall, Pedersen, and Cowling conductivities and corresponding currents.

Field-aligned currents flow along the magnetic field lines and they connect the ionosphere and the more distant regions of the magnetosphere.

Magnetotail currents are responsible for the long magnetic tails of planetary magnetospheres. Understanding the origin of magnetotail currents is important because they are tied to mechanisms that transfer the solar wind mass, momentum, and energy into planetary magnetic fields. The magnetotail currents are also a major source for auroral currents, since tail currents are diverted into the ionosphere along the magnetic field during aurora.

### 2.2 Definition of a current

The fundamental law that governs the behavior of currents is one of Maxwell's equations (Ampere's law), which relates the magnetic field **B** with the current density **J**:

$$\nabla \times \boldsymbol{B} = \mu_0 \left( \boldsymbol{J} + \epsilon_0 \frac{\partial \boldsymbol{E}}{\partial t} \right), \quad (1)$$

where $\mu_0$ is the permeability of free space, $\epsilon_0$ is the permittivity of free space and **E** is the electric field.

In the fluid description of plasma

$$\rho \frac{\mathrm{d}\boldsymbol{u}}{\mathrm{d}t} = -\nabla p + \boldsymbol{J} \times \boldsymbol{B}, \quad (2)$$

where $\rho$ is the mass density, **u** is the center mass velocity, and **p** is the plasma pressure.

In the magnetohydrodynamic (MHD) approximation the displacement current is ignored so that

$$\nabla \times \boldsymbol{B} = \mu_0 \boldsymbol{J}. \tag{3}$$

All currents in an MHD system must close on themselves ($\nabla \cdot \boldsymbol{J} = 0$). Currents in MHD fluids are coupled to the motion of fluids:

$$\boldsymbol{J} = \sigma(\boldsymbol{E} + \boldsymbol{u} \times \boldsymbol{B}), \tag{4}$$

where $\sigma$ is the electric conductivity.

Current density (the charge per second that flows across a unit area perpendicular to the flow direction) $\boldsymbol{J}$ is $\sum_s \boldsymbol{J}_s = \sum_s n_s q_s \boldsymbol{u}_s$, where $\boldsymbol{u}_s = \int d\boldsymbol{v}\ \boldsymbol{v} f_s(\boldsymbol{r},\boldsymbol{v},t) / \int d\boldsymbol{v}\ f_s(\boldsymbol{r},\boldsymbol{v},t)$ with the distribution function $f_s(\boldsymbol{r},\boldsymbol{v},t)$ of particle species s. The total current (the rate at which the charges are flowing out of the volume $V$ across the surface $S$ with normal $\boldsymbol{n}$) is $I = \sum_s I_s = \int \boldsymbol{J} \times \boldsymbol{n} dS$. Currents exist wherever there is a plasma.

## 2.3 Measurements of currents

### 2.3.1 Direct measurements of currents

Current density in space can be directly measured by particle detectors. The straightforward method to obtain the current density is to sum the average measured ion and electron current densities $\boldsymbol{j}_i = \langle n_i q_i \boldsymbol{v}_i \rangle$ and $\boldsymbol{j}_e = \langle n_e q_e \boldsymbol{v}_e \rangle$. However, it requires, as an input, complete information on the particles that are carrying the current. This in turn requires detecting all of the different particle species over all energies and pitch angles. Such measurements are extremely difficult to make because of the limitations of the detection technique. Moreover, the 3-D distribution function over all velocities must be measured.

There exists another practical problem, namely spacecraft charging. A spacecraft is usually several volts positive relative to the ambient plasma due to photoelectrons that are produced by the solar ultraviolet radiation interacting with the spacecraft. The photoelectron contribution must be separated from the naturally occurring current carries, whose energies may be similar to the energies of photoelectrons. Thus, the spacecraft charging distorts the measurement of low-energy particles.

This method was applied, for example, for the Geotail particle data (Frank et al., 1994) in the plasma sheet region ($-30\,R_E < x < -8\,R_E$, $|y| < 15\,R_E$) (Paterson et al., 1998; Kaufmann et al., 2001). The assumptions were that the current can be measured in thin current sheets with densities of $10\,\text{nA}\,\text{m}^{-2}$ or larger. Another assumption was that long-term averaging was necessary to reduce the effects of real fluctuations in the current density and flow velocity and influence of varying geomagnetic conditions from orbit to orbit. It was discovered that ions carry most of the cross-tail current on the duskside and that electrons carry most of the cross-tail current on the dawnside. Kaufmann et al. (2001) stress that it is very difficult to make measurements of current density with sufficient accuracy, especially for electrons. An example was given such that an error of $16\,\text{km}\,\text{s}^{-1}$ in the flow velocity will produce an error in the current density of $1\,\text{nA}\,\text{m}^{-2}$ when the density is $0.4\,\text{cm}^{-3}$. One of the conclusions was that direct current measurements are not accurate enough for development of a realistic magnetotail model.

### 2.3.2 Obtaining the perpendicular current from plasma pressure measurements

Another way to obtain the perpendicular current component is from the pressure gradient measurements. Under static conditions in the case of anisotropic plasma pressure, the current density $\boldsymbol{j}_\perp$ perpendicular to the magnetic field is given by Parker (1957)

$$\boldsymbol{j}_\perp = \frac{\boldsymbol{B}}{B^2} \times [\nabla p_\perp + (p_\parallel - p_\perp)\frac{(\boldsymbol{B} \cdot \nabla)\boldsymbol{B}}{B^2}], \tag{5}$$

where $p_\parallel$ and $p_\perp$ are plasma pressure parallel and perpendicular to the magnetic field, respectively. This equation is valid if a quasi-static equilibrium exists (force-balanced state) and there is no time dependence on the timescale of interest and inertial terms can be neglected. Other studies have also examined the relationship of plasma pressure to magnetic fields using this methodology (e.g., Heinemann and Pontius, 1990, 1991; Birmingham, 1992; Cheng, 1992; Heinemann et al., 1994).

The current is not directly measured but computed, and magnetic field and particle data are required to perform this calculation. If multi-satellite measurements are available, then plasma pressure gradients can, in principle, be computed. This requires a very exact cross calibration of the particle instruments on the different spacecraft. Moreover, it is necessary to subtract the spacecraft motion and to evaluate how the plasma structures are moving after this subtraction. It is also important that all spacecraft have a separation large enough to allow for a sufficient time drift to measure the structure's velocity ($\Delta t \leq t_{\text{spin}}$), but not so large as to violate the assumption of stationarity. This method is not accurate during very active periods. It is suitable for the perpendicular component only; the parallel component of the current cannot be calculated by pressure gradient estimate.

Figures 1 and 2 present the classical picture obtained using the measurements from AMPTE spacecraft radial profiles of the particle pressure perpendicular to the magnetic field and the computed current densities from four consecutive passes of AMPTE spacecraft during the 18–20 September 1984 storm (Lui et al., 1987).

This technique for obtaining the current densities was applied by using plasma pressure and magnetic field measurements from many different spacecraft during different activity periods. Zhang et al. (2011) used it for examining current carriers ahead of and within dipolarization fronts based on THEMIS measurements. They estimated the current density from ion bulk flow in the probe frame of reference, electron $E \times B$ drift current density, and electron pressure gradient current density.

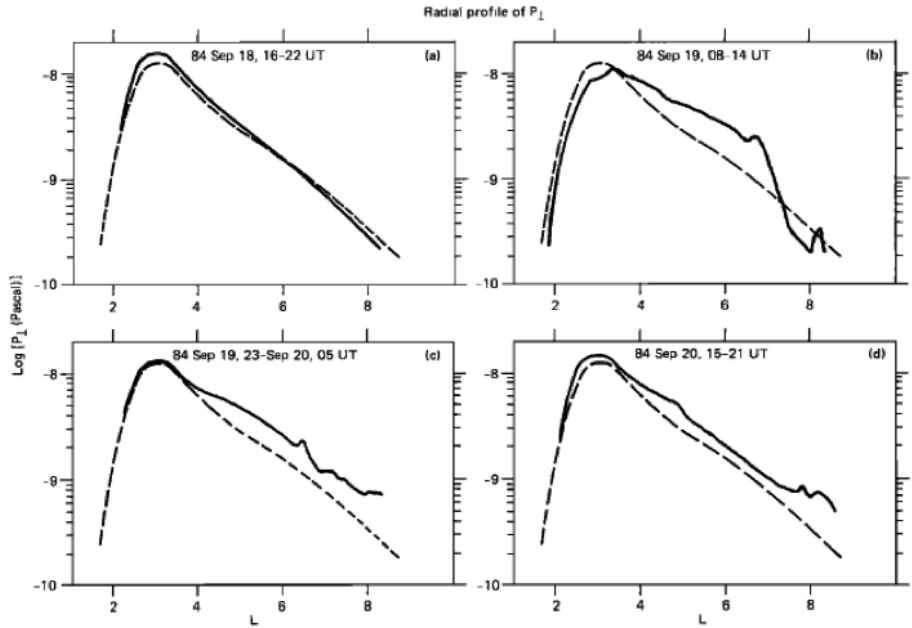

**Figure 1.** The radial profiles of the particle pressure perpendicular to the magnetic field from four consecutive passes of AMPTE spacecraft during the 18–20 September 1984 storm (Fig. 6 from Lui et al., 1987).

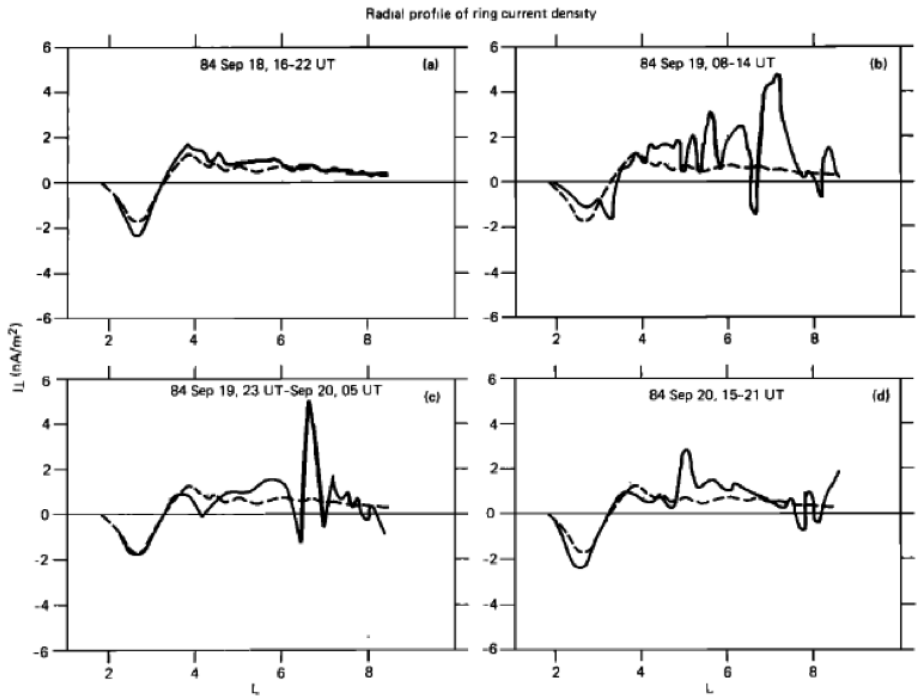

**Figure 2.** The radial profiles of the current densities for the four passes during the 18–20 September 1984 storm (Fig. 8 from Lui et al., 1987).

### 2.3.3 Computing electric currents using the plasma pressure extracted from ENA measurements

Electric currents can be also computed using the plasma pressure obtained from energetic neutral atom (ENA) measurements. The plasma pressure can be calculated from the global ion distributions extracted from the observed ENA images. If the plasma pressure $P$ distribution and the magnetic field $\boldsymbol{B}$ are known, the three-dimensional current system driven by the pressure gradients can be computed. This was done in Roelof et al. (2004), who used ENA images from the IMAGE HENA instrument that were inverted using a constrained linear inversion (DeMajistre et al., 2004) to obtain the proton distribution functions (Brandt et al., 2002b). It is possible then to compute the partial pressure over the energy range of ENA hydrogen measured by HENA. The perpendicular currents can be derived from the force-balance equation. It is also possible to calculate the field-aligned current flowing into the ionosphere according to Vasyliunas (1984). Roelof et al. (2004) computed the 3-D current system using an Euler potential formalism following Roelof (1989). A dipole magnetic field and an isotropic pressure were assumed. With an isotropic (scalar) pressure $P$, the current $\mathbf{J}$ is

$$\mathbf{J} = \nabla Q \times \nabla P \tag{6}$$

and $\nabla \times J = 0$. The second Euler potential $Q$ for a dipole field satisfies $B \times \nabla Q = 1$, so $Q$ is the partial volume of the flux tube ($Q = 0$ at the magnetic minimum-$B$ equator).

As was shown in Roelof et al. (2004) and Brandt et al. (2002a), the peak of the partial pressure occurs in the post-midnight sector.

### 2.3.4 Obtaining the current from magnetic field measurements

For steady-state currents, the contribution of displacement current can be ignored and Eq. (3) is valid. If $\nabla \times \boldsymbol{B}$ can be measured, it is possible to obtain the information on current density $\mathbf{J}$. However, just single point measurements in space do not give information on the gradients of the magnetic field components that are required to determine $\boldsymbol{B}$. Equation (3) can be applied to a quasi-one-dimensional current sheet configuration; the total current per unit length can be computed from single-spacecraft measurements above the current sheet as $I_y = 2B_x/\mu_0$ under the assumption that $B_x(z) = -B_x(-z)$, $|\,\mathrm{d}B_z/\mathrm{d}x\,| \ll |\,\mathrm{d}B_x/\mathrm{d}z\,|$ and the spacecraft is outside of the current sheet. This is also valid for a 1-D current disk with $B_r(z) = -B_r(-z)$, $|\,\mathrm{d}B_z/\mathrm{d}r\,| \ll |\,\mathrm{d}B_r/\mathrm{d}z\,|$. Although only total current density (per unit length) can be estimated from this method, the current density (per unit area) can be estimated if the spacecraft moves fast across a quasi-stationary 1-D current sheet. In such a case, the current density profile across a current sheet can be reconstructed. Such configurations often occur when a low-altitude satellite crosses the field-aligned current sheet (e.g., Dubyagin et al., 2003) or the magnetopause moves with respect to the spacecraft (e.g., Anekallu et al., 2011).

If multi-spacecraft data are available, a curlometer technique can be used. This technique was successfully applied using data obtained simultaneously on board the four Cluster spacecraft (Escoubet et al., 2001). Measurement accuracy of the current density in such approach can be substantially affected by

- the tetrahedral geometry of the four spacecraft,
- the size (in time and space) of the current structure sampled,
- the linear interpolation made between various measurement points,
- the eventual experimental errors inherent to the magnetometer.

The Cluster mission is based on four identical spacecraft launched on similar elliptical polar orbits with a perigee at about 4 $R_\mathrm{E}$ and an apogee at 19.6 $R_\mathrm{E}$ (Escoubet et al., 2001). The inter-spacecraft separation strategy was planned in order to allow for the study of the various plasma structures encountered by Cluster along the orbit (Chanteur and Mottez, 1993; Robert et al., 1998). The maneuvers to change the inter-spacecraft separation took place once or twice a year, depending on the spatial scales of the plasma structures to be studied. The tetrahedron formed by the four spacecraft can thus have characteristic sizes ranging between 100 km and a few Earth radii. On board each spacecraft, 11 experiments permit a wide variety of measurements of the plasma parameters (particles and fields). Among the instruments on board is a fluxgate magnetometer (FGM).

The curlometer technique has been described in detail by Dunlop et al. (1988, 2002). Taking into account Eq. (3), in a discrete Cartesian coordinate system,

$$(\nabla \times \boldsymbol{B})_x \approx \frac{\Delta B_z}{\Delta y} - \frac{\Delta B_y}{\Delta z}, \tag{7}$$

$$(\nabla \times \boldsymbol{B})_y \approx \frac{\Delta B_x}{\Delta z} - \frac{\Delta B_z}{\Delta x}, \tag{8}$$

$$(\nabla \times \boldsymbol{B})_z \approx \frac{\Delta B_y}{\Delta x} - \frac{\Delta B_x}{\Delta y}. \tag{9}$$

These equations can be applied to the Cluster data (four simultaneous points of the magnetic field measurements) to evaluate the magnetic field gradients (over the spacecraft) and thus the current density through the tetrahedron formed by the four spacecraft.

The main assumptions for curlometer technique are as follows (Vallat et al., 2005):

- stationarity in the region of interest, assuming the field does not vary on timescales of the spacecraft motion;
- the field varies slowly and linearly inside the tetrahedron;
- all measurement points are situated inside the same current sheet, which implies that the current density is constant inside the tetrahedron.

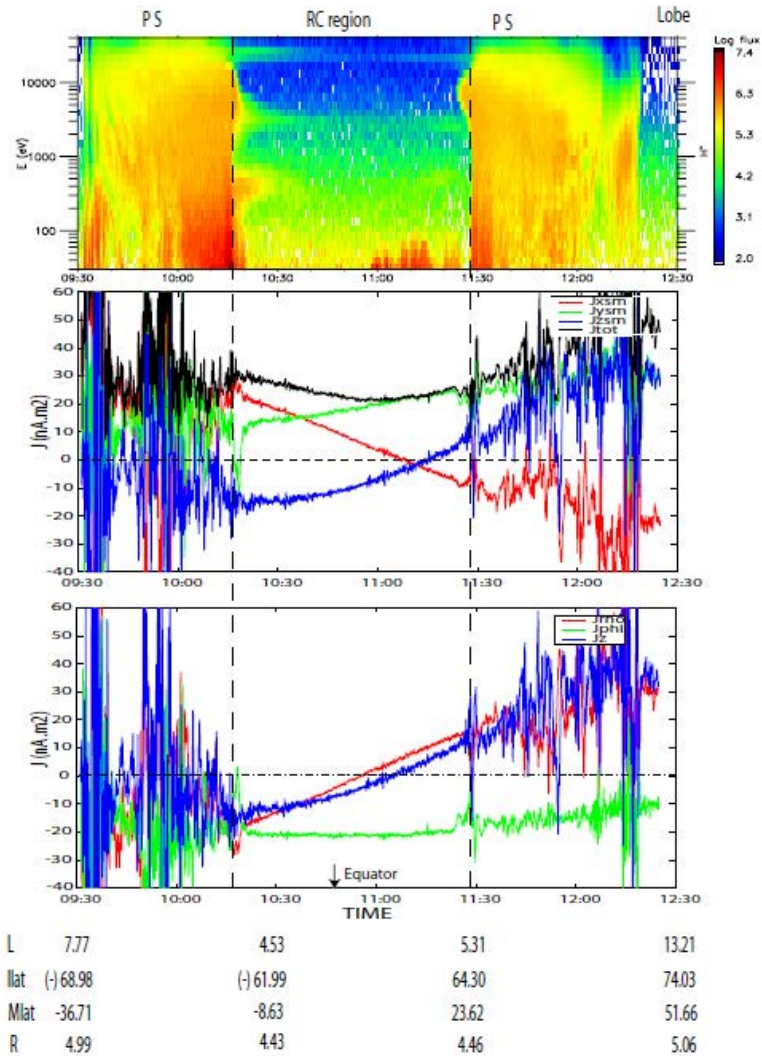

**Figure 3.** Cluster data for the 18 March 2002 event: H+ energy–time spectrogram for SC4 in particle flux units (ions cm$^{-2}$ sr s keV), current density components in the SM coordinate system and in nA m$^{-2}$ (second panel), and in the local cylindrical coordinate system (bottom panel). Black dashed lines demarcate the ring current region. L shell, invariant latitudes, magnetic latitudes and geocentric distances are indicated below (Fig. 9 from Vallat et al., 2005).

Figure 3 demonstrates the results of the curlometer technique for 18 March 2002 event.

The Active Magnetosphere and Planetary Electrodynamics Experiment (AMPERE) exploits observations of magnetic perturbations from the nearly 70 polar-orbiting Iridium satellites to reconstruct the field-aligned current patterns above the northern and southern polar ionospheres at 10 min cadence (Anderson et al., 2000; Waters et al., 2001; Green et al., 2006). The magnetic perturbations from each satellite are used to constrain a spherical harmonic fit of the global magnetic perturbation pattern, the curl of which provides the field-aligned current pattern.

### 2.4 Drifts and currents

It is important to note the critical distinction between particle guiding-center drifts and electric currents in the magnetospheric plasmas. The electric current density, an observable quantity, consists in an average over the particle distribution (not guiding center). The guiding-center drifts are a time average of the motion of a single particle. These two terms are not equivalent (in fact they can be quite different) and therefore referring to the "ring current carried by westward-drifting ions" is technically incorrect. As it is shown below, the portion of the current corresponding to the westward ion guiding-center drift is completely canceled by part of the magnetization current, with the remainder of the magnetization current, proportional to the plasma pressure gradient, being the sole quantity that dictates the direction and magnitude of the transverse electric current. The electric current is the same in the fluid and particle pictures, but care must be taken to conduct the calculation to the appropriate level of detail in order to obtain the same result. That is, assumptions made within the derivation process can lead to erroneously different equations for the cross-magnetic field current. In the fluid picture (quasi-static), the plasma momentum equation (simplified for isotropic pressure) takes the simple form (see, for example, Rossi and Olbert, 1970, Chapter 9):

$$\mathbf{J} = \frac{\mathbf{B}}{B^2} \times \nabla P. \tag{10}$$

The same value for the current in a quasi-static configuration can be obtained from the particle picture, but again, care must be taken to include all the relevant terms. It is easy to show this for a particular case with a certain field orientation, without the results losing any of their general validity (Goldston and Rutherford, 1995). For the general case, the electric current "carried" by the guiding-center drift is

$$\begin{aligned}\mathbf{J}_G &= (n_i \langle W_{\perp i} \rangle + n_e \langle W_{\perp e} \rangle) \mathbf{B} \times \nabla B / B^3 \\ &= [n \langle W_\perp \rangle (\mathbf{B} \times \nabla B)]/B^3, \end{aligned} \tag{11}$$

where $\langle W_\perp \rangle$ is the total perpendicular energy (ions and electrons combined).

The magnetic moment due to plasma magnetization is

$$\mathbf{M} = -\frac{n \langle W_\perp \rangle}{B^2} \mathbf{B}, \tag{12}$$

corresponding to a current density

$$\mathbf{J}_M = \nabla \times \mathbf{M} = -\nabla \times \left( \frac{n \langle W_\perp \rangle}{B^2} \mathbf{B} \right). \tag{13}$$

Equation (13) is a curl of the product of a scalar and a vector, so the result is the sum of the scalar times the curl of the vector and the gradient of the scalar crossed with the vector. That is, there are two physical scenarios contributing to a magnetization current: the presence of a local shear or twist in the magnetic field, or the presence of a local pressure gradient. In the first case, if there is a current flowing in the region, then the plasma in that region will have a magnetization current that cancels part or all of that other current. The second case means that a localized peak in pressure will have a current flowing around it.

Adding the two current densities for the special configuration with the magnetic field $B_x = 0$, $B_y = 0$, $B_z = B_z(y)$ (current in the $x$ direction), one obtains

$$\begin{aligned}\mathbf{J}_x &= (\mathbf{J}_G + \mathbf{J}_M) \cdot \mathbf{x} \\ &= -\frac{n \langle W_\perp \rangle}{B^2} \frac{dB}{dy} - \frac{d}{dy}\left[ \frac{n \langle W_\perp \rangle}{B} \right] \\ &= -\frac{1}{B}\frac{d}{dy}(n \langle W_\perp \rangle), \end{aligned} \tag{14}$$

equivalent to $\mathbf{J} = \frac{\mathbf{B} \times \nabla P}{B^2}$, which is exactly the result obtained from the fluid picture above. We have shown therefore that the guiding center contributes nothing to the total current density, the only contribution arising from a term in the magnetization current proportional to the pressure gradient. Let us reiterate this point: in the isotropic case, the only transverse electric current is from the magnetization term and is flowing around the pressure peak. However, this current is not constant, as $\mathbf{J}$ is also inversely proportional to $\mathbf{B}$. If the field intensity changes across the plasma pressure peak, then the transverse current density will also change intensity. Closure of this unbalanced perpendicular current must then be carried by field-aligned and ionospheric currents.

The case where the field includes a curvature is similar, with the formulation for the current density $\mathbf{J}_C$ being nearly identical to Eq. (11) except that the perpendicular energy $W_\perp$ is replaced by 2 times the parallel energy $W_\parallel$.

$$\mathbf{J}_C = \left[2n\langle W_\parallel \rangle (\mathbf{B} \times \nabla B)\right]/B^3 \tag{15}$$

Summing them all together and accounting for parallel and perpendicular plasma pressures, the total transverse current then becomes as in Eq. (5). Only in the case of anisotropic pressure is there a contribution to perpendicular current density from the curvature term. For example, in a hypothetical situation, in a uniform pressure plasma in a non-uniform magnetic field the net current density is zero, even though the particles still drift.

## 3 Definitions: permanent current systems

### 3.1 Westward symmetric ring current

The symmetric ring current is one of the oldest concepts in magnetospheric physics. A current of a ring shape flowing around the Earth was first introduced by Stormer (1907) and supported by Schmidt (1917). Chapman and Ferraro (1931, 1941) used a ring current concept for the model of a geomagnetic storm. Studies by Lui et al. (1987), Spence et al. (1989) and Lui and Hamilton (1992) obtained the radial plasma pressure profiles in the midnight magnetosphere with pressure increasing earthward with a peak around 3 $R_E$ and then decreasing toward the Earth (see Figs. 1 and 2). This plasma pressure profile corresponds to a two-part ring current, with westward current outside of the pressure peak and eastward current inside of the pressure peak. It was also found that this structure exists for all times. Quiet-time ring current can be $\sim$ 1–4 nA m$^{-2}$ and storm-time ring current can reach $\sim$ 7 nA m$^{-2}$. From the observational point of view, the ring current is never purely symmetric (Jorgensen et al., 2004; Le et al., 2004). It can be more symmetric during quiet times but during storm times it is asymmetric, especially in the main and early recovery phases. One of the most comprehensive analyses was done by Le et al. (2004), in which 20 years of magnetospheric magnetic field data from ISEE, AMPTE/CCE and Polar missions were examined. Le et al. (2004) used the intercalibrated magnetic field data, con-structed the statistical magnetic field maps and derived 3-D current densities. Figure 4 summarizes their analysis, showing the derived equatorial ring current intensity as a function of magnetic local time and distance from the dipole axis for four Dst* intervals. The ring current has for a long time been considered as a measure of the ground disturbance of the magnetic field (e.g., Akasofu and Chapman, 1961; Kamide and Fukushima, 1971; Kamide, 1974). The averaged magnetic field depression observed at low latitude is used to derive the Dst index (Sugiura, 1964). The local time asymmetry of the ground magnetic can be used to measure the asymmetry of the ring current (Akasofu and Chapman, 1964; Cahill, 1966).

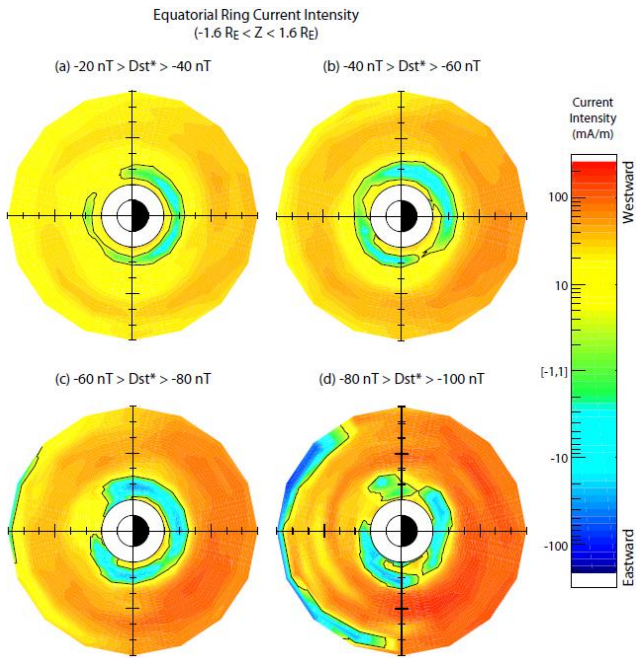

**Figure 4.** Equatorial ring current intensity as a function of magnetic local time and distance from the dipole axis for each of the four Dst* levels (Fig. 8 from Le et al., 2004).

There have been numerous in situ observations of the ring current, including particle measurements providing plasma pressure and current estimated from it (Frank, 1967; Smith and Hoffman, 1973; Lui et al., 1987; Spence et al., 1989; De Michelis et al., 1997; Milillo et al., 2003; Korth et al., 2000; Ebihara et al., 2002; Lui, 2003), as well as deriving the current from the measurements of the magnetic field in the inner magnetosphere (Le et al., 2004; Vallat et al., 2005; Ohtani et al., 2007). Remote sensing of ENAs emitted from the ring current conveyed the global information about the ring current morphology, dynamics and composition starting from ISEE-1 spacecraft observations (Roelof, 1987) and continued in IMAGE and TWINS missions (Pollock et al., 2001; Mitchell et al., 2003; Brandt et al., 2002a; Buzulukova et al., 2010; Goldstein et al., 2012).

Several studies have investigated the nonlinear inflation of the magnetic field due to a symmetric westward ring current in the inner magnetosphere. The work of Dessler and Parker (1959) and Sckopke (1966) yielded a formula relating the total energy content of the plasma within the Earth's dipole field to the globally averaged ground-based magnetic perturbation (the Dessler–Parker–Sckopke, or DPS, relation). Carovillano and Siscoe (1973) demonstrated that the DPS relation applies even if the plasma pressure is not azimuthally symmetric about the Earth. Liemohn (2003) went on to show that the plasma pressure must drop to zero inside the integration domain, as otherwise the DPS relation includes a truncation current effectively dropping the pressure to zero at the outer boundary. This, in a rough sense, approximates the contribution of currents beyond the integration domain and helps justify the usage of the DPS relation for spatially limited drift physics model results (e.g., Jordanova et al., 1996; Liemohn et al., 1999). Similarly, Ganushkina et al. (2012) showed that the DPS relation does not match direct Biot–Savart integration when the plasma distribution is flowing through a nondipolar magnetic field.

Some studies have contemplated the extreme limits of magnetic field distortion in the presence of a very intense symmetric westward ring of current. Parker and Stewart (1967) examined the nonlinear effects of a plasma pressure peak inflating the dipole, showing that extreme conditions and magnetic topologies can arise when the total energy content of the plasma approaches that of the magnetic field. Sozou and Windle (1969) showed that, if an embedded current were large enough, magnetic nulls could be created within the inner magnetosphere. A similar result was found by Lackner (1970) using a Vlasov kinetic treatment of the plasma instead of a guiding-center approach. Vasyliunas (2011) explored the question of the largest possible current that could be supported within Earth's inner magnetosphere, determining it could reach a Dst value of $-2500$ nT. Vasyliunas (2013) followed up by estimating that superstorm-level solar driving conditions could reach this level in a timescale as short as 2–6 h.

### 3.2 Eastward symmetric ring current

As was mentioned above, the AMPTE/CCE data (Lui et al., 1987; Lui and Hamilton, 1992; De Michelis et al., 1997) revealed the plasma pressure profiles corresponding to a two-part ring current. The derived current densities for the eastward ring current were typically about $2\,\mathrm{nA\,m^{-2}}$ for quiet and storm times with the most of the current carried between 2 and 3 $R_E$ (see Fig. 2).

The existence of this current system was confirmed by later observational studies. De Michelis et al. (1997) produced the average ring current patterns based on particle measurements from the AMPTE/CCE CHEM instrument for four different local time sectors. They found both the eastward and the westward components of the ring current. At the same time, Nakabe et al. (1997) estimated the current structure from visual inspection of the magnetic field maps obtained from DE-1 magnetic field data. In their analysis, the intense (up to 50 nA m$^{-2}$) eastward current was only evident on the dayside, which is in contradiction to previous studies. Jorgensen et al. (2004) analyzed the magnetic field data from the CRRES satellite by spatial location and produced magnetic field maps to calculate then the local current systems by taking the curl of the magnetic field. They found an eastward-directed component to the ring current, as well as a westward-directed component, and these two currents were consistent with the plasma pressure peak located at approximately 3.5 $R_E$.

Several studies have been performed to explicitly include the eastward ring current into the magnetospheric magnetic field modeling. The first attempt was made by Lui et al. (1994). This current system was missing from the existing global magnetospheric magnetic field models, including first versions of the Tsyganenko models like T87 (Tsyganenko, 1987) and T89 (Tsyganenko, 1989). It was demonstrated that although this eastward ring current may not change the magnetic field significantly, its absence considerably affects the plasma pressure distribution required to maintain equilibrium with the magnetic forces. Eastward current was included in later versions of global (Tsyganenko, 2000, 2002) and event-oriented (Ganushkina et al., 2004) magnetic field models.

## 3.3 Asymmetric and partial ring current with closure by the region 2 field-aligned currents

The concept of the partial ring current and its closure to the ionosphere was suggested by Alfvén in the 1950s (Egeland and Burke, 2012). According to the review paper by Egeland and Burke (2012), Alfvén distinguished between the gyro-motion and the guiding-center motion and showed field-aligned currents coupled to the auroral ionosphere. The general form of the perpendicular current density is given by Eq. (5) (Parker, 1957). This equation implies that **J** is distributed symmetrically when **B** and *P* are distributed symmetrically about the center of the Earth. Such a situation would rarely or never occur in the magnetosphere, because the magnetosphere is essentially asymmetric, compressed by the solar wind dynamic pressure on the dayside, and stretched by the tail current on the nightside.

In addition to that, the plasma pressure distribution during disturbed times becomes highly asymmetric due to plasma transport and injection from the nightside plasma sheet to the inner magnetosphere. The resulting plasma distribution presents a gradient in the azimuthal direction resulting in the spatial asymmetry of the ring current (Roelof, 1987; Ganushkina et al., 2000; Liemohn et al., 2001). In the inner magnetosphere, the plasma pressure (i.e., energy density) is primarily due to ions. About 90 % of the energy density comes from ions with energy less than a few hundreds of keV (Williams, 1981; Daglis et al., 1993). Contributions from high-energy ions with energies of up to 4 MeV have also been suggested (Lui and Hamilton, 1992). In situ satellite observations by Viking show a strong asymmetry between the dusk- and dawnside ion distribution after the onset of a magnetic storm. Korth et al. (2000) showed that the enhancement of the ions occurs on the nightside and duskside first, followed by the dawnside. Based on statistical studies of particle data, the energy density is distributed fairly symmetrically during geomagnetically quiet times (Ebihara et al., 2002; Lui, 2003), whereas it becomes asymmetric during high-AE (De Michelis et al., 1999), low-Dst (Ebihara et al., 2002), and high-Kp (Lui, 2003) periods. Ebihara et al. (2002) found that the pressure (or the energy density) becomes asymmetric during the storm main phase, whereas it becomes symmetric during the recovery phase. Statistical studies of the magnetic field have also shown that the degree of the asymmetry becomes large for low Dst (Terada et al., 1998; Le et al., 2004).

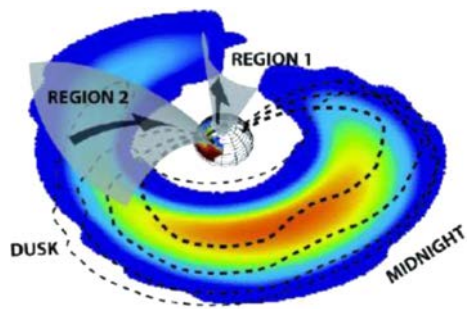

**Figure 5.** Current systems associated with the partial ring current as deduced from the ENA measurements (Plate 2 from Brandt et al., 2008).

Temporal variation in the plasma pressure was successfully captured by the ENAs observations. ENAs are emitted by a charge-exchange collision between energetic ions and neutrals. After reconstruction of the three-dimensional pressure distribution, current systems related to the high-pressure region were obtained by Roelof (1989), Roelof et al. (2004), Roelof and Skinner (2000), and Brandt et al. (2004, 2008). Figure 5 shows the current systems associated with the partial ring current, indicating that the partial ring current is connected to the region 2-sense field-aligned current.

In general, the perpendicular current cannot be closed in the inner magnetosphere. The remnant of the perpendicular current must flow along a field line to complete a closure of the current (Vasyliunas, 1970; Wolf, 1970). Additional electric fields are established to conduct away the space charge deposited by the field-aligned current in the ionosphere. In the steady-state condition, the Pedersen current is responsible for closing the current between the field-aligned current. In the case shown in Fig. 5, the eastward electric field is established to close the pair of the region 2-sense field-aligned current, and the northward electric field is established to close the region 1 and region 2 field-aligned currents. The first one is called the shielding electric field, and is observable by the ground-based magnetometer and radars when the convection electric field almost vanishes (Fejer et al., 1979; Kelley et al., 1979; Spiro et al., 1988; Kikuchi et al., 2008; Ebihara et al., 2008). Such conditions are called overshielding. At the same time, in most of the cases it is merely a reduction of the westward electric field, and the predominant part of region 2 is closed with region 1. The latter electric field may be related to the westward, rapid plasma flow observed in the sub-auroral region (Galperin et al., 1973; Spiro et al., 1979). The sub-auroral westward flow is observed to be temporally variable during a storm time, which is probably a manifestation of the complex structure of the plasma pressure in the inner magnetosphere (Ebihara et al., 2009).

### 3.4 Tail current with closure via return current on magnetopause

The discovery that the Earth's magnetotail extended beyond the Moon's orbit came as a surprise in the mid-1960s. Using in situ magnetic field observations by the IMP-1 satellite, Ness (1965) and Speiser and Ness (1967) showed that the nightside geomagnetic field trailed out far behind the Earth in the antisolar direction forming the magnetotail. Unlike the ring current, whose existence was predicted well before the space era, the finding of the thin sheet of the equatorial current concentrated near the magnetic field reversal region and dividing the magnetotail into two slab-like regions with almost uniform magnetic field of opposite direction surprised many. This picture was confirmed soon by direct observation of the equatorial plasma sheet (e.g., Singer et al., 1965; Anderson and Ness, 1966; Bame et al., 1967). These first results were followed by the extensive exploration of the system's geometry (e.g., Russell and Brody, 1967; Fairfield, 1979, 1980; Slavin et al., 1985; Owen et al., 1995; Tsyganenko et al., 1998), plasma population (e.g., Frank, 1967) and dynamics (e.g., Behannon and Ness, 1966; Fairfield and Ness, 1970; Aubry and McPherron, 1971; Hones et al., 1971; McPherron, 1972; Hones et al., 1973; Kokubun and McPherron, 1981). The special importance of the cross-tail current sheet comes from the fact that it is a locus of the instabilities leading to the magnetospheric substorm (Hones, 1979; Lui, 1991; Baker et al., 1996).

Although the cross-tail current and the ring current (symmetric and partial) are considered to be separate current systems, there is no evidence of any discontinuity between these two currents on the nightside. Sugiura (1972) claimed that the current in the inner magnetosphere is a continuation of the tail current sheet. Apparently, the current continuously passes from the cross-tail current in the magnetotail into the ring current in the inner magnetosphere. On the other hand, obviously, the near-Earth ring current and the far tail current occupy the regions characterized by different particle drift paths (trapped and open, respectively; Alfvén, 1955; Harel et al., 1981a, b), different trajectories of the thermal particles in the equatorial region (adiabatic and chaotic, respectively, e.g., Chen, 1992; Delcourt et al., 1996), and different anisotropy of the pressure tensor (dominance of the perpendicular pressure and almost isotropic pressure, respectively) (e.g., Stiles et al., 1978; De Michelis et al., 1999). Although on average, the cross-tail current can be considered a diamagnetic current carried by thermal protons (in the stationary magnetospheric frame), the physics can be much more complex for the extremely thin current sheets (e.g., Asano et al., 2005; Artemyev et al., 2010; Nakamura et al., 2006; Runov et al., 2006) or during the bursty bulk flows which are ubiquitous in the magnetotail (e.g., Baumjohann et al., 1990; Angelopoulos et al., 1994; Runov et al., 2011). The tail current responds to the interplanetary magnetic field (IMF) much faster than the ring current does (Tsyganenko, 2000; Tsyganenko and Sitnov, 2005) and it can be used as another way to distinguish it from the ring current. During a substorm dipolarization, injected plasma has an associated partial ring current with it between the tail and pre-existing ring currents.

The various definitions of the tail and ring currents were debuted in studies of the contribution of the different current systems to the Dst index during geomagnetic storms (e.g., Dremukhina et al., 1999; Ohtani et al., 2001; Maltsev, 2004; Ganushkina et al., 2004; Kalegaev et al., 2005). Some of these studies were motivated by the study of Iyemori and Rao (1996), who showed that Dst slowed down its drop during the substorm onset (substorm current wedge (SCW) development). Since it is believed that the tail current is diverted to ionosphere during a substorm (McPherron, 1972), these authors implicitly define the tail current as a westward equatorial current exactly at and outside the region of the current disruption. However, the current disruption models do not give a strict definition of the tail current. They mostly only assume that it is a thin sheet current (e.g., Pulkkinen et al., 1994). Alexeev et al. (1996) determined the inner edge of his model's tail current as the equatorial projection of the maximum of the midnight auroral electrojet along the dipole field at $r = 4$–$7\,R_E$ depending on the magnetospheric activity (Alexeev et al., 1996; Dremukhina et al., 1999). The majority of studies of the current disruption have been conducted using the geosynchronous spacecraft. For this reason, many authors define the tail current as a current outside the $r = 6.6\,R_E$ (e.g., Ohtani et al., 2001). Turner et al. (2000) estimated the contribution to the Dst of the equatorial current in the $X = [-50\,R_E, -6\,R_E]$, $Z = [-5\,R_E, 5\,R_E]$ box. Other definitions have also been used. Maltsev et al. (1996) and Maltsev (2004) defined the tail current as the current outside the $B = B_m$, where $B_m$ is the magnetic field magnitude at the subsolar point of the magnetopause. This definition is based on the simple assumption that 90° pitch-angle particles drift along $B = $ const curve, an assumption which is true only for zero electric field. However, the authors mentioned that the tail current defined in such a way also includes partial ring current (Maltsev, 2004). Skoug et al. (2003) distinguished the current flowing in the $y$ direction as opposed to the cur-rent carried by the particle undergoing the azimuthal drift. Recently, Liemohn et al. (2015) compiled statistics of numerical model results for all of the major storms of solar cycle 23, finding that the timing and intensity of near-Earth nightside current undergoes a systematic progression through a storm sequence, with the tail current dominating in the early main phase.

Since the first observations, it was realized that the cross-tail current had to close over the magnetopause, forming a theta-like system (Axford et al., 1965). However, it is not that simple to answer the question of which part of the magnetopause current should be referred to as the tail current. The family of the Tsyganenko models (Tsyganenko, 1995; Tsyganenko and Sitnov, 2005) employs the separate module for every current system. The field of a module/system comprises the field of the system itself and the field of the shielding magnetopause current so that the normal component of the total module field on the magnetopause is zero. When the contributions of the different systems are summed, the surface magnetopause currents of the different systems may cancel each other out. Figure 5 of Tsyganenko and Sibeck (1994) shows that the shielding currents of the cross-tail current on the dayside magnetopause flow westward, in the opposite direction to the currents shielding Earth's dipole. Thus, if one traces the streamlines of the tail current module alone, some of them come to the dayside magnetopause, but it would not be the case if the shielding currents of the all other systems were taken into account. It is questionable whether the current system shielding current should be considered an inherent part of every system or the magnetopause currents shielding the magnetic fields of all current systems should be considered a separate current system. As regards the latter case, it should be noted that there is no way to define the magnetopause return current of the cross-tail current uniquely and separately from the shielding magnetopause currents.

Finally, there several possible/existing definitions of the cross-tail current: (1) nightside equatorial westward current outside 6.6 $R_E$; (2) westward equatorial current closing on the magnetopause; (3) current which flows in the $y$ direction, in contrast to circular/azimuthal ring current; (4) westward equatorial current outside the inner edge of the (electron/ion) plasma sheet (Alfvén zero-energy layer); (5) westward current in the region of the stretched magnetic field; (6) westward current in the region of isotropic plasma pressure; (7) in the region of the quasi-one-dimensional magnetic configuration; (8) a current carried by $< 20$ keV particles; (9) the westward equatorial current exactly at and outside the region of the current disruption during the substorm; and (10) westward equatorial current directly driven by southward IMF component.

Figure 6 shows current traces from the Space Weather Modeling Framework (SWMF) (Toth et al., 2005) under idealized input conditions of steady driving with IMF $B_Z = -5$ nT. The SWMF configuration for this simulation included the Block-Adaptive-Tree Roe-type Solar wind Upwind Scheme (BATS-R-US) MHD model (Powell et al., 1999) for the global magnetospheric solution, the Ridley Ionosphere Model (Ridley et al., 2004) for the ionospheric electrodynamics solution, and the Rice Convection Model (Jaggi and Wolf, 1973) for capturing the inner magnetospheric drift physics processes. This figure was made from the same simulation as presented in Fig. 9 of Liemohn et al. (2013a), but this new figure is from a much farther vantage point in order to focus on the entire magnetosphere rather than just the near-Earth nightside. The plots shows the total current density in the equatorial and meridional plans and the field-aligned current density on the 2.5 $R_E$ radius sphere, which is the inner boundary of the MHD model in this simulation. The magnetopause current is clearly visible, along with a less intense bow shock current in front of it and a near-Earth nightside current behind the Earth. The current traces, extracted from the MHD results, are colored to categorize them into the various current systems. The green lines show the Chapman–Ferraro current loops flows across the dayside magnetopause and closing behind the cusps. The pink lines are the region 1 field-aligned current system, closing just inside the magnetopause over the pole. Farther back in the $-x$ direction is the tail current, encircling the tail lobes as it flows across the equatorial plane and then closing along the magnetopause. The final current system shown in Fig. 6 is the region 2 partial ring current loop as black lines, closing through the ionosphere just equatorward of the oppositely directed region 1 current system. This figure is meant as a numerically derived schematic diagram of the canonical locations of the various current systems relative to each other.

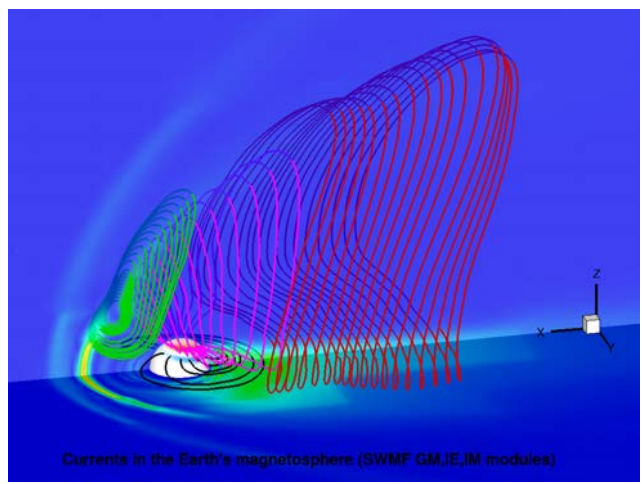

**Figure 6.** Current lines obtained by global magnetospheric modeling with the SWMF. The view is from near dusk with the Sun to the left, the inner white sphere is has a 2.5 $R_E$ radius, the color on the sphere shows field-aligned current intensity, and the background color shows total current density in the $y = 0$ and $z = 0$ planes. The colors of the lines represent different current systems: green is the Chapman–Ferraro magnetopause current, pink is the region 1 field-aligned current system, black is the region 2 and partial ring current system, and red is the tail current.

## 3.5 Chapman–Ferraro magnetopause current

Sydney Chapman and Vincenzo Ferraro were the first to explain the basic nature of the interaction between the solar wind and the Earth's magnetic field in the 1930s (Chapman and Ferraro, 1931). They suggested that the magnetosphere carves out a cavity in the solar wind and that neither solar wind plasma nor the solar wind magnetic flux has access to this cavity. The thin boundary that separates the magnetosphere from the solar wind is a current sheet, known as the magnetopause. However, it was not until the early 1960s that first measurements of this boundary were made, by Explorer 10 and 12, confirming the theory of Chapman and Ferraro (e.g., Cahill and Amazeen, 1963).

When the solar wind interacts with the magnetic field of the Earth, a shock front forms in front of the magnetosphere, the bow shock, which acts to slow down the solar wind so that plasma can flow around the magnetosphere. As the solar wind passes through the shock, it is decelerated, heated, and diverted around the Earth in a region called the magnetosheath. This region has a thickness of about 3 $R_E$ near the sub-solar point but increases rapidly in the downstream direction. After being decelerated by the bow shock, the heated solar wind plasma is accelerated again from subsonic to supersonic flow as it moves around the flanks of the magnetosphere.

The magnetopause separates the plasma of the magnetosheath, in which particle pressure plays the major role, from the more tenuous magnetospheric plasma, in which magnetic pressure is dominant. The magnetic field inside the magnetosphere points roughly northward, whereas the orientation of the magnetic field in the magnetosheath is determined by the clock angle of the interplanetary magnetic field. Hence, the magnetopause marks the location where the magnetic field changes both in strength and direction, and as a consequence, an extensive current flows across the magnetopause. In the simplest picture, as magnetosheath protons and electrons enter the higher magnetic field inside the magnetosphere, they perform a half-gyration and re-enter the sheath. As protons and electrons gyrate in opposite directions around the magnetospheric field, their differential motion within the boundary produces a current. The magnetic field gradient effectively provides a magnetic pressure that excludes particles from the magnetosphere, appearing as the $\bm{j} \times \bm{B}$ term in the plasma momentum equation. The direction of current flow is determined by the orientation of the magnetic field within the boundary, resulting in dawn-to-dusk current across the nose of the magnetosphere and dusk-to-dawn flow across the high-latitude magnetopause tailwards of the cusp openings. As indicated in Fig. 6 by the green current circuits, the magnetopause or Chapman–Ferraro currents form closed loops across the sunward-facing surface of the magnetosphere, with an average current density of 20 mA m$^{-1}$. The thickness of the current layer is related to the ion gyroradius of the (heated) magnetosheath ions in the magnetospheric field, on the order of several hundred kilometers.

To first approximation, the magnetic field strength in the magnetosheath is low. The effect of the magnetopause current is to produce a magnetic perturbation that cancels the dipole of the Earth outside the boundary. This necessarily produces a doubling of the undisturbed dipole field strength just inside of the magnetopause (e.g., Chapman and Bartels, 1940). The magnetopause forms where the magnetic pressure associated with the doubled ("compressed") dipole magnetic field counteracts the thermal pressure of the magnetosheath. In equilibrium, the magnetic pressure inside the magnetopause $P_{\text{mag}} = (2B)^2/2\mu_0$, where $B = B_{\text{eq}}(R_E/R_{\text{MP}})^3$ is the dipole magnetic field strength at the location of the magnetopause, $R_{\text{MP}}$, and $B_{\text{eq}}$ is the equatorial magnetic field strength of the Earth, equal to the sum of thermal and magnetic pressures in the magnetosheath, which, in turn, is equal to the dynamic or ram pressure of the solar wind. Hence, the location and strength of the magnetopause currents at the nose of the magnetosphere are dependent on the solar wind dynamic pressure: $P_{\text{ram}} = \rho v^2$. Away from the nose, the current magnitude decreases as the magnetopause is further from the Earth, where the magnetic field is weaker. The lower magnetic pressure is compensated for as the magnetopause is no longer perpendicular to the Sun–Earth line and the pressure exerted by the solar wind/magnetosheath is lower. The shape of the magnetopause, also known as tail flaring, is dictated by this balance (e.g., Coroniti and Kennel, 1972; Petrinec and Russell, 1996). Shue et al. (1998), using ISEE 1 and 2, AMPTE/IRM and IMP 8 measurements, constructed an empirical model that calculates the magnetopause standoff distance as well as level of tail flaring based on solar wind velocity and density.

Under normal solar wind conditions, the subsolar magnetopause is located approximately 10 $R_E$ upstream of the Earth. When the dynamic pressure of the solar wind increases the magnetopause current intensifies and moves closer to the Earth. The magnetic perturbation due to the current can be sensed at the surface of the Earth. As this perturbation is of opposite polarity to the perturbation associated with the ring current, Dst can display a positive "initial phase" excursion associated with the solar wind shock that precedes the "main phase" of a geomagnetic storm.

Under extreme solar wind driving associated with strongly southward IMF, it is postulated that the region 1 current merges with the Chapman–Ferraro current on the dayside magnetopause and that it is the region 1 current that largely stands off the solar wind (Siscoe et al., 2002). It is proposed that this limits the current that can flow in the region 1 field-aligned current circuit, and consequently limits the cross-polar cap potential associated with magnetospheric convec-tion, in a phenomenon known as transpolar voltage saturation.

Note that the current in the sense of Chapman–Ferraro current can be carried by energetic particles orbiting the magnetic minimum of the cusp region (e.g., Erlandson et al., 1988; Chen et al., 1998). Niehof et al. (2008, 2010) further quantified the relationship of cusp diamagnetic cavities with the presence of cusp energetic particles. It is clear that some of the current, especially that flowing close to or even within the cusp funnel structure, is carried by MeV-energy particles. Although the cusp currents (inside the cusp funnel) can be considered as a natural continuation of the Chapman–Ferraro magnetopause currents, the background physical conditions are rather different. Chapman–Ferraro currents mostly flow in a region of strong flow shear (gradient) and strong $\nabla B$. For cusp currents inside the funnel there is no flow and $\nabla B$ is moderate. Also, cusp currents are less dependent on IMF orientation.

### 3.6 Region 1 field-aligned currents

Kristian Birkeland, after whom the current system is named, first proposed the existence of currents (Birkeland, 1908) flowing parallel to the magnetic field to help explain magnetic disturbances observed in the polar regions. He also undertook terrella experiments to confirm his predictions. In the subsequent years, several authors proposed theories on the physical properties and generation of Birkeland currents, including Alfvén (1950), Martyn (1951), Fejer (1961), Swift (1965) and Cole (1963). It was not until the space age that its physical existence was confirmed by Boström (1967) and Cummings and Dessler (1967) using magnetic field data from the low-altitude, polar-orbiting Triad satellite (Zmuda et al., 1966, 1967). Zmuda and Armstrong (1974), also using Triad magnetometer data, showed that this current system consisted of oppositely directed, but closely spaced in latitude, concentric sheets. Iijima and Potemra (1976a) first cataloged the Birkeland current system into region 1 and region 2 currents, where region 1 currents were defined as currents directed toward the Earth on the dawnside and upward on the duskside and region 2 currents defined as opposite in sign and lying equatorward of the region 1 system. Figure 7 shows a summary plot of the current system taken from Iijima and Potemra (1976a). Further from the Earth, evidence of the existence of the Birkeland current systems was found using magnetic field measurements by Aubry et al. (1972), Fairfield and Ness (1972), Sugiura (1975) and Ohtani et al. (1988). Later, by using models or assumptions about the ionospheric conductance, Birkeland current distribution maps were also deduced from radar measurements (Sofko et al., 1995) and data assimilation methods such as AMIE (Richmond and Kamide, 1988; Lu et al., 1996). More recently, detailed maps of Birkeland currents in the ionosphere have been produced from the magnetometers on the Iridium satellite constellation (Anderson et al., 2000, 2005), which can then be used to produce a picture of the time evolution of the currents (Wilder et al., 2012; Clausen et al., 2012, 2013; Coxon et al., 2014a, b). Even with the caveats associated with the techniques used to derive the Birkeland currents, these maps suggest that the current system is a lot more structured and dynamic than the statistically derived current patterns of Iijima and Potemra (1976b) as shown in Fig. 7.

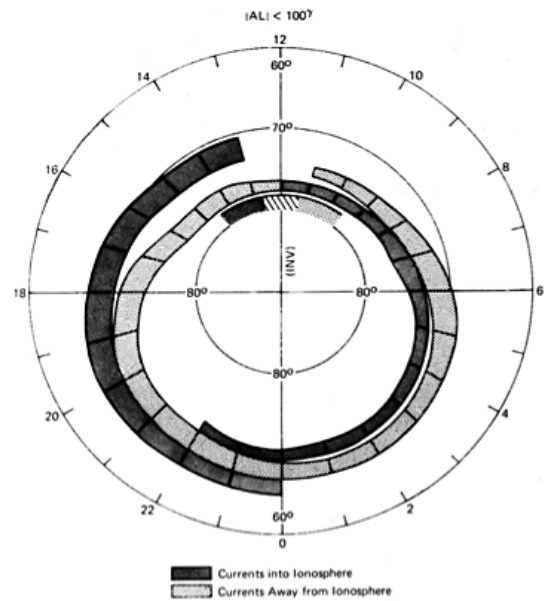

**Figure 7.** A pattern of the distribution of large-scale field-aligned currents (Fig. 6 from Iijima and Potemra, 1976b) determined from TRIAD data for weakly disturbed conditions. The "hatched" area in the polar cusp region corresponds to unclear current flow directions.

Further development of the Iridium into the AMPERE (Anderson et al., 2014) has provided a new and global view of field-aligned currents on a relatively high temporal cadence. With satellites on six orbital planes, AMPERE is able to produce a field-aligned current map every 10 min. This product is allowing a new examination of R1 currents and their response to solar wind driving (Korth et al., 2014). Merkin et al. (2013), for instance, compared AMPERE with MHD simulations of an interplanetary shock arrival at Earth to investigate the timing and intensity of current system changes, demonstrating that both techniques provide reasonable estimates of the total current but perhaps miss small-scale peaks.

The Birkeland current system is thought to be an indicator of the coupling of plasma processes in the magnetosphere to the ionosphere (e.g., Siscoe et al., 1991). However, the exact physical processes associated with the formation of region 1 currents are still unclear, and are believed to depend strongly on whether the associated magnetic field line is an open (connects to the solar wind) or closed (connects to the other hemisphere) field line. Evidence for region 1 currents residing on open field lines was presented, for example, by Cowley (2000); Yamauchi et al. (1993), while evidence for region 1 currents residing on closed field lines was suggested by Saflekos et al. (1979) and D'Angelo (1980). Region 1 field-aligned currents on both open and closed field lines were found by Ohtani et al. (1995), Xu and Kivelson (1994), Haraguchi et al. (2004) and Wang et al. (2008). Region 1 currents that reside on open field lines are believed to be driven by the solar wind, which acts as a generator (Iijima and Potemra, 1982; Stern, 1983; Siscoe et al., 1991) possibly by dayside reconnection. On closed field lines, their formation can be due to processes taking place in either the boundary layer (Lotko et al., 1987) or in the plasma sheet (e.g., Antonova and Ganushkina, 1997; Wing and Newell, 2000; Toffoletto et al., 2001). To further complicate things, another example of a region 1 Birkeland current system on closed field lines is the so-called substorm current wedge (SCW) (McPherron et al., 1973a) that appears during substorms (Clausen et al., 2013). Global MHD simulations by Raeder et al. (2001) and more recent local simulations by Yang et al. (2012) reproduce the SCW. In the work by Yang et al. (2012), the formation of the SCW was attributed the current system to the injection of low-content flux tubes into the inner magnetosphere during a substorm expansion.

One of the dangers when looking at current systems is the temptation to interpret current systems as having physical meaning like wires in a circuit and that the current is the cause of the magnetic field. While this interpretation may be convenient for constructing magnetic field models, care must be taken in giving physical causality to the currents. As pointed out by Vasyliunas (2005) for space plasmas, this interpretation is not the case: "Over the wide range of timescales from electron plasma period to Alfvén wave travel time, there simply is no way to calculate the changing currents except by taking the curl of the changing magnetic fields; statements about changes of current are not explanations but merely descriptions of changes in the magnetic field." That is, Vasyliunas (2005) found that the currents can be considered as a diagnostic, not a cause of the magnetic field, so the interpretations of the region 1 currents are then to be understood simply as a product of the coupling between the magnetosphere and the ionosphere.

### 3.7 Region 2 field-aligned currents

As discussed above in the section on the asymmetric and partial ring current, during disturbed times the ring current is asymmetric and a partial ring current develops, driven by the plasma pressure gradients in the inner nightside magnetosphere. This partial ring current closure is through the ionosphere, and a field-aligned current system develops connecting the westward partial ring current to the auroral electrojet. This is the region 2 field-aligned current system (R2 FAC), which is just equatorwards of the R1 FAC system that connects the cross-tail current to the ionosphere. The large-scale field-aligned current system organization, in terms of region 2, region 1 and region 0 currents, was initially determined by Iijima and Potemra (1976a) through analysis of the Triad satellite magnetometer measurements. Whereas the ring current is a current perpendicular to the magnetic field and its density $\boldsymbol{j}$ is given by the MHD equations – i.e., it is directly related to the perpendicular pressure gradient (see also section on the asymmetric and partial ring current) – the field-aligned current density $j_\parallel$ has to be calculated from the divergence of the current (Vasyliunas, 1970):

$$\nabla \times (\boldsymbol{j}_\perp + \boldsymbol{j}_\parallel) = 0. \tag{16}$$

The field-aligned current density is then given by the equation

$$B \frac{\partial}{\partial s}\left(\frac{j_\parallel}{B}\right) = 2 \times \boldsymbol{j}_\perp \times \frac{\nabla B}{B} + \left(\frac{\boldsymbol{B}}{B^2}\right) \times \nabla \\ \times \left[\left(\frac{\boldsymbol{B}}{B^2}\right) \times \left(\rho_m \frac{\mathrm{d}\boldsymbol{U}}{\mathrm{d}t} + \nabla \times \boldsymbol{p}\right)\right], \tag{17}$$

where

$$\frac{\partial}{\partial s} = \left(\frac{\boldsymbol{B}}{B}\right) \times \nabla \tag{18}$$

is the gradient operator along the direction of the magnetic field, $\rho_m$ is the charged particle mass density and $\boldsymbol{U}$ is the fluid velocity. Note that the original work of Vasyliunas (1970), which only considered an isotropic plasma distribution, was extended by Birmingham (1992), who derived an equation for $j_\parallel$ in the presence of an anisotropic plasma. A schematic of the R2 FAC system, connecting the partial ring current to the ionosphere, is given in Fig. 5. The current lines, in this schematic, have been deduced from the plasma pressure distribution calculated from ENA images (Roelof et al., 2004; Brandt et al., 2008). Magnetic local time (MLT)–magnetic latitude maps of the R2 FAC system have been obtained from the analysis of the Iridium magnetometer data (Roelof et al., 2004), whereas the relationship between the R2 FAC system and the ring current for this figure has been modeled by Zheng et al. (2006).

### 3.8 R0 and NBZ dayside field-aligned currents

Iijima and Potemra (1976b) found, using TRIAD magnetometer data, that in the midday sector there is often another large-scale FAC system on the poleward side of the R1 system. They referred to this FAC system as cusp current because of the proximity of its location to the magnetic cusp. Later, Erlandson et al. (1988) compared the latitudinal structures of FACs and particle precipitation measured by the Viking satellite and concluded that it is not the "cusp" current but the midday R1 current that is collocated with cusp-related soft particle precipitation; they referred to such a R1 current as the traditional cusp current. Bythrow et al. (1988) conducted a similar study but with DMSP data and found that the "cusp" current is actually collocated with mantle precipitation, which is characterized by soft ion precipitation with its energy decreasing poleward. Later, however, Ohtani et al. (1995) reported that in general the boundaries of large-scale FACs do not coincide with the boundaries of particle precipitation (in other words, there is no one-to-one correspondence between FACs and particle precipitation) and therefore they simply referred to this most poleward dayside FAC as the (midday) R0 current. The term R0 current has been used before, for example, by Heikkila (1984) to refer to a current system that possibly surrounds the pole on the poleward side of the R1 system.

One of the most important characteristics of the R0 system is that its spatial distribution strongly depends on the IMF $B_y$ component. In the Northern Hemisphere, the R0 current flows predominantly out of the ionosphere for positive IMF $B_y$ and into the ionosphere for negative IMF $B_y$ (e.g., Wilhjelm et al., 1978; Iijima et al., 1978; Papitashvili et al., 2002). This pattern may be envisioned in such a way that the demarcation of the dawnside and duskside R0 currents shifts to postnoon and prenoon for positive and negative IMF $B_y$, respectively, in the Northern Hemisphere. The situation is opposite in the Southern Hemisphere (Erlandson et al., 1988). The R0 current appears to be always paired with the R1 current even if its distribution is skewed significantly by IMF $B_y$. This strongly suggests that, in the midday sector, the R0 and R1 currents are associated with the zonal convection (Wilhjelm et al., 1978), which is presumably driven by dayside reconnection that takes place off the noon meridian depending on the IMF orientation.

The northward IMF $B_z$ (NBZ) system is also distributed poleward of the R1 system, but it is morphologically different from the R0 system. Whereas the R0 current sheet is oriented zonally forming a pair with a R1 current adjacently equatorward, the NBZ current is distributed inside the polar cap. The NBZ current flows into and out of the ionosphere on the dusk- and dawnsides, respectively. The current system is named after the fact that it appears during strongly northward IMF $B_z$ (Iijima et al., 1984). Presumably the NBZ system is related to the sunward convection that takes place in the middle of the polar cap during northward IMF $B_z$ (Maezawa, 1976), which, along with two conventional convection cells farther equatorward, may be envisioned as a four-cell convection pattern (Reiff and Burch, 1985). The high-latitude reconnection, the reconnection between the IMF and the lobe magnetic field, is the most likely cause. Figure 8 presents the sketch of the dayside Birkeland currents from Erlandson et al. (1988).

### 3.9 Ionospheric currents

Currents flowing in geospace, including the magnetopause and the ring current, produce magnetic perturbations that can be detected at the ground. In 1859, Richard Carrington was the first to realize that the ultimate cause of these perturba-tions was disturbances on the Sun; since that time it has been the goal of solar–terrestrial science to understand the chain of events that transmits solar disturbances to near-Earth space to produce the magnetic perturbations that are observed. A key element of the chain is currents flowing horizontally in the ionosphere at altitudes of 100–130 km, providing closure for currents flowing up and down magnetic field lines from their generator in the magnetosphere. The most important of these field-aligned currents are the region 1 currents flowing near the open–closed field line boundary (or polar cap boundary), mapping to a generator on the magnetopause, and the region 2 currents which close the circuit through a partial ring current in the inner magnetosphere.

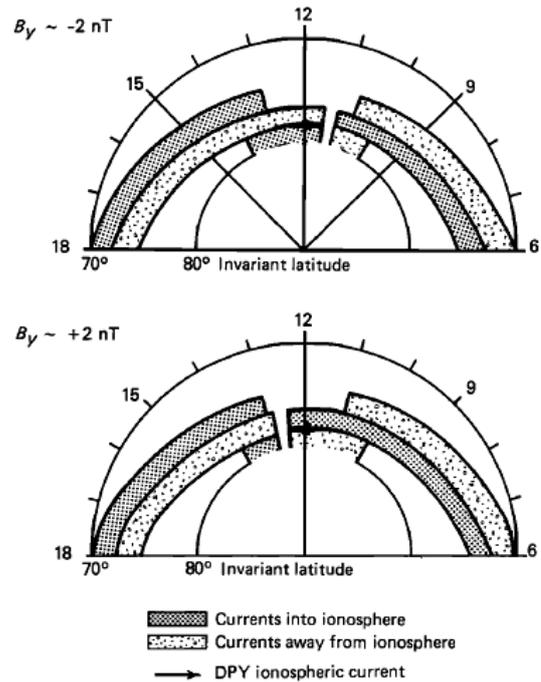

**Figure 8.** Sketch of the dayside Birkeland currents, modified from the statistical pattern developed by Iijima and Potemra (1976b), in the Northern Hemisphere for IMF $B_y$ near $-2$ nT and $+2$ nT (Fig. 11 from Erlandson et al., 1988).

Electric fields imposed from the magnetosphere above are associated with flow of plasma in the ionosphere (see Fig. 9). In the collisionless regime, at ionospheric altitudes above 150 km (F region), charged particles gyrate around the magnetic field direction (clockwise for electrons when looking along the magnetic field direction and counterclockwise for ions) and drift horizontally with velocity $\boldsymbol{E} \times \boldsymbol{B}/B^2$; ions and electrons drift together and no net current flows. This motion can be thought of as the motion of plasma frozen to the magnetic field as it circulates with the Dungey cycle of magnetospheric convection (Dungey, 1961). Below the F region the atmospheric density increases and collisions between charged and neutral particles become increasingly frequent as altitude decreases. The ion–neutral collision fre-quency exceeds the electron–neutral collision frequency, so ions are more collisionally (or frictionally) bound to the neutrals at any given altitude.

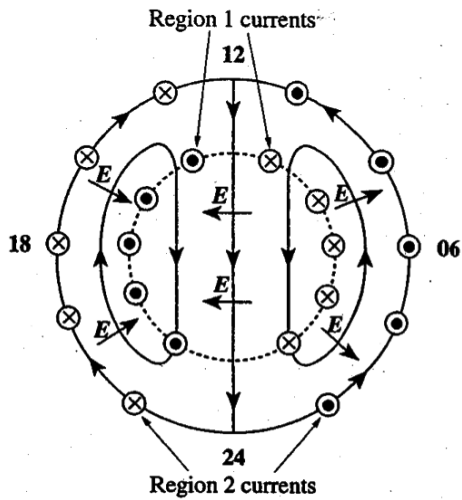

**Figure 9.** Schematic of the ionospheric flow streamlines (arrowed curves) and associated electric field pattern. Arrows facing inward and outward show the locations of field-aligned currents, the inner and outer rings being the region 1 and region 2 currents, respectively. The dashed line marks the open–closed field line boundary (Fig. 1 from Cowley, 2000).

Collisions have the effect of bringing the ions and electrons momentarily to rest, imparting momentum to the neutrals and heating them. The charged particles are thereafter accelerated, ions in the direction of $E$ and electrons in the direction of $E$, before continuing to $E \times B$ drift. This differential acceleration results in a separation of the drift velocities of ions and electrons – that is, horizontal current flow. The magnitude and direction of the current depends on the ratios of the electron and ion gyro-frequencies $\Omega_e$ and $\Omega_i$ and the electron–neutral and ion–neutral collision frequencies $\nu_{en}$ and $\nu_{in}$: as $\nu$ approaches $\Omega$, particles perform fewer gyrations after each collision, and the bulk drift of ions rotates from the $E \times B$ direction towards $E$, and towards $-E$ for electrons. In addition, as $\nu$ becomes very significant, the motions of particles becomes increasingly impeded, and little current flows. At the top of the collisional region, the ion bulk speed is somewhat reduced and rotated from $E \times B$ towards $E$; relative to the $E \times B$-drifting electrons, this results in a current with components in the $E$ and $-E \times B$ directions. As altitude decreases, the ion drift slows and rotates further towards the $E$ direction, while electrons rotate towards the $-E$ direction and the current flow becomes increasing directed along $E$. The currents flowing in the directions parallel to $E$ and to $-E \times B$ are known as Pedersen and Hall currents, respectively, and the dependence of the current densities (A m$^{-2}$) on the strength of the driving electric field $E$ is given by the Pedersen and Hall conductivities, $\sigma_P$ and $\sigma_H$, such that, in its simplest form (ignoring lesser contributions from $\nu_{en}$ and $\Omega_e$),

$$j = \frac{ne(\nu_{in}/\Omega_i)}{1+(\nu_{in}/\Omega_i)^2}\left[\frac{E}{B} - (\nu_{in}/\Omega i)\frac{E \times B}{B^2}\right]$$
$$= \sigma_P E + \sigma_H \hat{B} \times E,$$

where $e$ is the electronic charge and $n$ is the electron density – the density of charge carriers (Maeda, 1977). The interplay between the differing directions of ion and electron flow and the speeds of their drift as a function of altitude results in Pedersen and Hall conductivities maximizing at altitudes of 125 and 110 km, respectively. The height-integrated conductivities are known as the Pedersen and Hall conductances, $\Sigma_P$ and $\Sigma_H$.

In terms of the global pattern of currents, the Pedersen currents act to close upward and downward field-aligned currents, mainly the region 1 and region 2 currents, while Hall currents flow in the direction opposite to the flow streamlines of the ionospheric convection pattern. In a uniform conductance ionosphere, the Hall current is divergence-free, whereas divergence of the Pedersen conductance occurs at the FAC regions. Gradients in the conductances, associated with gradients in $n$, can lead to further divergences in the Hall and Pedersen currents, which must also be closed by field-aligned currents. Such gradients occur between the day- and nightside ionospheres due to the gradient in photoionization, and between the auroral zone, where $n$ is increased by impact ionization associated with particle precipitation, and the polar cap or sub-auroral ionosphere. Region 1 currents are stronger than region 2 currents as Pedersen closure current can flow over the polar cap as well as through the return flow region, though typically the current magnitude is lower in the polar cap due to the lower conductance. The relationship between the ionospheric Pedersen currents, the ionospheric conductance, and the field-aligned region 1 and region 2 currents has been exploited in a simple analytical model (Milan et al., 2013) to predict the current magnitudes based on the expanding/contracting polar cap paradigm, a time-dependent version of the Dungey cycle (Dungey, 1961; Milan et al., 2007). The predictions of this model are largely borne out by observations from AMPERE (Clausen et al., 2012, 2013; Coxon et al., 2014a, b).

While all currents produce a magnetic perturbation, the perturbations associated with the high-latitude closed loops of upward/downward current and Pedersen closure current largely cancel (Fukushima, 1976), so the main magnetic deflection measured on the ground is associated with Hall currents. These Hall currents are strongest in the auroral zones due to enhanced conductivity and are directed eastwards in the dusk sector and westwards in the dawn sector, commonly known as the eastward and westward electrojets, sometimes shortened to eastjet and westjet, and also known collectively as the DP2 current system. The northward (southward) magnetic deflection produced by the eastjet (westjet) is measured by the AU (AL) index (Davis and Sugiura, 1966). Dur-ing non-substorm intervals, the AU and AL indices provide an indicator of the combined effect of convection strength and conductances in the auroral zone. During substorms, the westjet is supplemented by westward current associated with the substorm current wedge in the night sector (also known as the DP1 current system), and AL is enhanced.

Seasonal variations in photoionization result in unequal conductances in the summer and winter hemispheres. On the other hand, observations show that the large-scale electric field associated with magnetospheric convection is broadly equal in both hemispheres (e.g., de la Beaujardiere et al., 1991). As a consequence, the ionospheric currents driven in the summer and winter hemispheres, and hence the region 1 and region 2 currents that feed them, are unequal (e.g., Ridley, 2007). On a smaller scale, in regions where the conductivity is greatly enhanced, such as within the substorm auroral bulge, the frictional coupling between charged particles and the neutral atmosphere can become sufficient to fix magnetic flux in the ionosphere, forming a barrier to convection (e.g., Kirkwood et al., 1988; Morelli et al., 1995). This is described as "line tying" or it is said that the convection electric field is shorted out in this region. Convection can proceed outside of the high-conductivity region but must flow around the barrier. Convection can only fully resume once the conductivity has diminished.

In the case of electric fields imposed from above, the horizontal currents transfer momentum from the solar wind dynamo via the region 1 currents to the ionosphere and atmosphere. Momentum sources in the ionosphere, i.e., neutral winds, can push charged particles across the magnetic field, generating electric fields and hence currents. The major ionospheric dynamo-generated current system is formed by the large-scale thermosphere prevailing wind and tide pattern produced by differential solar heating, manifesting as counter-rotating horizontal current vortices in the Northern and Southern Hemisphere dayside ionospheres, known as the solar quiet (Sq) system (Kato, 1956). A much weaker lunar-cycle-driven current system also exists. At polar latitudes, motions of the neutral ionosphere driven by momentum transfer during prolonged periods of intense magnetospheric convection can persist after driving has ceased, producing a "flywheel effect" current which can couple back up to the magnetosphere.

## 4 Definitions: temporal current systems

### 4.1 Substorm current wedge (SCW) with R1 FAC closure

The substorm current wedge (SCW) with the downward field-aligned current on its dawnside and the upward current on the duskside is the main current system responsible for the strong magnetic field disturbance during the magnetospheric substorms. The SCW was proposed as a 3-D closure current of the substorm westward electrojet from around the mid-

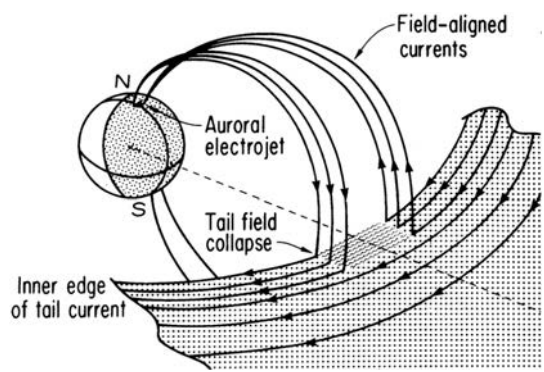

**Figure 10.** Simple line current model of substorm expansion with perspective view of a diversion of the inner edge of the tail current (Fig. 7 from Clauer and McPherron, 1974).

1960s and thereafter (Boström, 1964; Atkinson, 1967; Akasofu and Meng, 1969; Meng and Akasofu, 1969; Bonnevier et al., 1970; Rostoker et al., 1970; Kamide and Fukushima, 1972; Crooker and McPherron, 1972) and the system has become widely accepted after seminal papers by McPherron et al. (1973a, b). Figure 10 presents a simple line current model of SCW from Clauer and McPherron (1974). Now the SCW is thought as a deviation of the disrupted equatorial current to the ionosphere during the substorm.

It was shown that the location of its upward and downward current roughly coincide with the west and east terminations of the auroral bulge, respectively (e.g., Untiedt et al., 1978; Baumjohann et al., 1981; Gelpi et al., 1987; Sergeev et al., 1996). Like auroral bulge, the SCW is not a static structure; once it is formed it broadens azimuthally and radially (Nagai, 1982; Lopez and Lui, 1990; Jacquey et al., 1993; Ohtani et al., 1998). Although the sense of the SCW system for a typical substorm which initiates at the midnight–premidnight local time sector is the same as for the large-scale region 1 system, there are a few important differences. First, the magnetospheric part of the SCW is located in the inner magnetosphere at $r < 15\,R_E$ (Takahashi et al., 1987; Lui et al., 1988; Lopez et al., 1989; Jacquey et al., 1993; Lu, 2000), obviously not the same region where the current of the region 1 flows. Second, the substorm onsets are observed at various local times within the 19:00–02:00 MLT sector (Liou et al., 2001; Frey et al., 2004). Since the initial substorm electrojet intensification may occur in the narrow longitudinal sector (Bonnevier et al., 1970; Opgenoorth et al., 1980), the SCW can be localized entirely on the dusk- or dawnside at least during the initial stage of the expansion phase.

At present, there is no consensus on the physical mechanism of the SCW formation. Two competing scenarios are fast flow braking (Baker et al., 1996; Shiokawa et al., 1997, 1998) and cross-tail current disruption initiated by current driven instability (Lui, 1991, 1996). The former mechanism is supported by the results of MHD modeling of the fast flow propagation (Birn and Hesse, 1991; Scholer and Otto, 1991;

Birn and Hesse, 1996). The observations have shown that the vorticity of the bulk velocity (Keiling et al., 2009) and azimuthal plasma pressure gradient (Yao et al., 2012) both contribute to the SCW field-aligned currents' generation.

From the very beginning, it was clear that SCW is a very simplified model of the real current system of the substorm. Observations (Ohtani et al., 1990; Sergeev et al., 2011) and modeling (Birn et al., 1999; Yang et al., 2012; Birn and Hesse, 2013) have shown that the real system consists of multiple wedges of the different sense, scale and intensity. Although the typical SCW system is dominant, at least during the initial part of the expansion phase, the intensity of the secondary wedges can also be significant. Moreover, the simulation results (Birn et al., 1999, 2004) imply that the SCW-like system may be formed every time the burs bulk flow comes to the inner magnetosphere region, even if it does not lead to the substorm development. This complex picture makes defining the SCW system a difficult task. The commonly accepted definition is that SCW is the current system developing during substorm main phase as a result of the deviation of the cross-tail current to the ionosphere where it closes via westward electrojet.

### 4.2 Asymmetric Birkeland currents into the conjugate hemispheres

Østgaard and Laundal (2012) summarized findings from conjugate auroral imaging and, combined with earlier theoretical studies, suggested three mechanisms that can produce interhemispheric or asymmetric currents and different auroral brightness in the two hemispheres. The relevance and importance of these mechanisms have been the subject of several studies. Here we review some of these recent results about two of these mechanisms.

One mechanism that can lead to hemispheric differences in region 1 field-aligned currents is due to hemispheric differences in the solar wind dynamo efficiency when the IMF has a significant $B_x$ component. According to the open magnetospheric model (Dungey, 1961), magnetic flux is opened on the dayside and closed on the nightside. The proposed mechanism describes a current generator on the magnetopause. As the opened magnetic flux tubes are draped tailward, the tension force on these flux tubes will tend to slow them down and give rise to a current on the magnetopause. Lobe flux tubes are also open flux tubes and will also have a tension force acting on them, but further down the tail. As stated in Reistad et al. (2014), the model work by Siscoe et al. (2000) shows that this magnetopause current can close as region 1 current in the ionosphere. As first noticed by Cowley (1981), the orientation of the IMF in the $xz$ plane would lead to different strength of the tension force in the two hemispheres, as shown in Fig. 11a. This tension force gives rise to a current generator, and as parts of these currents close in the ionosphere (Fig. 11b), interhemispheric differences in auroral brightness should be seen in the dusk sector.

Laundal and Østgaard (2009) reported a significantly brighter aurora in the southern dusk that lasted for more than an hour. With a $B_x > 0$-dominant IMF, this observation is consistent with this mechanism. Other support for this mechanism can be found in Shue et al. (2002), which reported an overall brighter aurora in the Northern Hemisphere for IMF $B_x < 0$. Reistad et al. (2013) investigated 19 h of simultaneous global conjugate auroral data containing 10 sequences with duration from 1 to 5 h during active geomagnetic conditions. They identified 15 features of non-conjugate aurora, meaning features that were only observed in one hemisphere or a feature that was significantly more intense in one hemisphere compared to the other. They found that seven of these features were consistent with the solar wind dynamo mechanism.

Following these results, Reistad et al. (2014) explored whether the difference in solar wind dynamo efficiency can have a statistically significant impact on the aurora. In their study, the entire IMAGE WIC data set was used. Careful selection criteria were implemented to avoid the effect of other possible mechanisms. See Reistad et al. (2014) for details. The results are shown in Fig. 11c–h. In the Northern Hemisphere the superposed images (Fig. 11c and d) are comprised of more than 150 observations in the MLT sector from 17 to 24, while for the Southern Hemisphere images (Fig. 11f and 11G) there are more than 80 observations in the same MLT sector. As can be seen in Fig. 11e and 11h, there are distinct intensity differences between the negative and positive IMF $B_x$ cases. The differences are seen in the dusk sector (15:00–19:00 MLT in the north and 16:00–20:00 MLT in the south) and at the poleward edge, most clearly in the Northern Hemisphere. This is exactly as expected from the efficiency difference of the solar wind dynamo due to IMF $B_x$ component where the upward region 1 current closes in the poleward region of the ionospheric dusk sector. A Kolmogorov–Smirnov test (see Reistad et al., 2014, their Figs. 2e and 3e) showed that the differences are significant at the 95 % confidence level within most of the indicated regions.

Another mechanism pointed out by Østgaard and Laundal (2012) is related to the $B_y$ component of the IMF. They referred to the explanation suggested by Stenbaek-Nielsen and Otto (1997), which was based on earlier observations of a non-uniform $B_y$ component in the closed magnetosphere. These earlier observations suggested that $B_y$ has a gradient towards the Earth in the tail region and, due to Ampere's law, they argued that this gradient gives rise to an interhemispheric current. Although this description is consistent with observations of non-conjugate aurora from a conjugate aircraft campaign (Stenbaek-Nielsen and Otto, 1997), it does not provide a detailed description of how the asymmetric stresses in the tail can propagate from the common generator region in the equatorial plane to the ionosphere(s).

Here we will suggest a modified scenario of how IMF $B_y$ induces a $B_y$ component in the closed magnetosphere, and argue that the result is not an interhemispheric current but rather an asymmetric current from the plasma sheet into the two hemispheres. First, merging with the Earth's magnetic field (both during subsolar and lobe reconnection) will result in a dawn–dusk asymmetry of the open magnetic flux in the lobes in the two hemispheres. This is shown in Fig. 12a for positive IMF $B_y$ and is the same as Fig. 3a in Liou and Newell (2010). This effect will be opposite in the two hemispheres, and consequently the forces acting on the field lines in the two hemispheres will be oppositely directed (Cowley, 1981; Liou and Newell, 2010). These asymmetric magnetic pressure distributions forced by the IMF will also affect closed field lines and control the longitudinal asymmetry of the foot points. The result is an added $B_y$ component in the closed magnetosphere in the same direction as the IMF $B_y$, as seen in Fig. 12a.

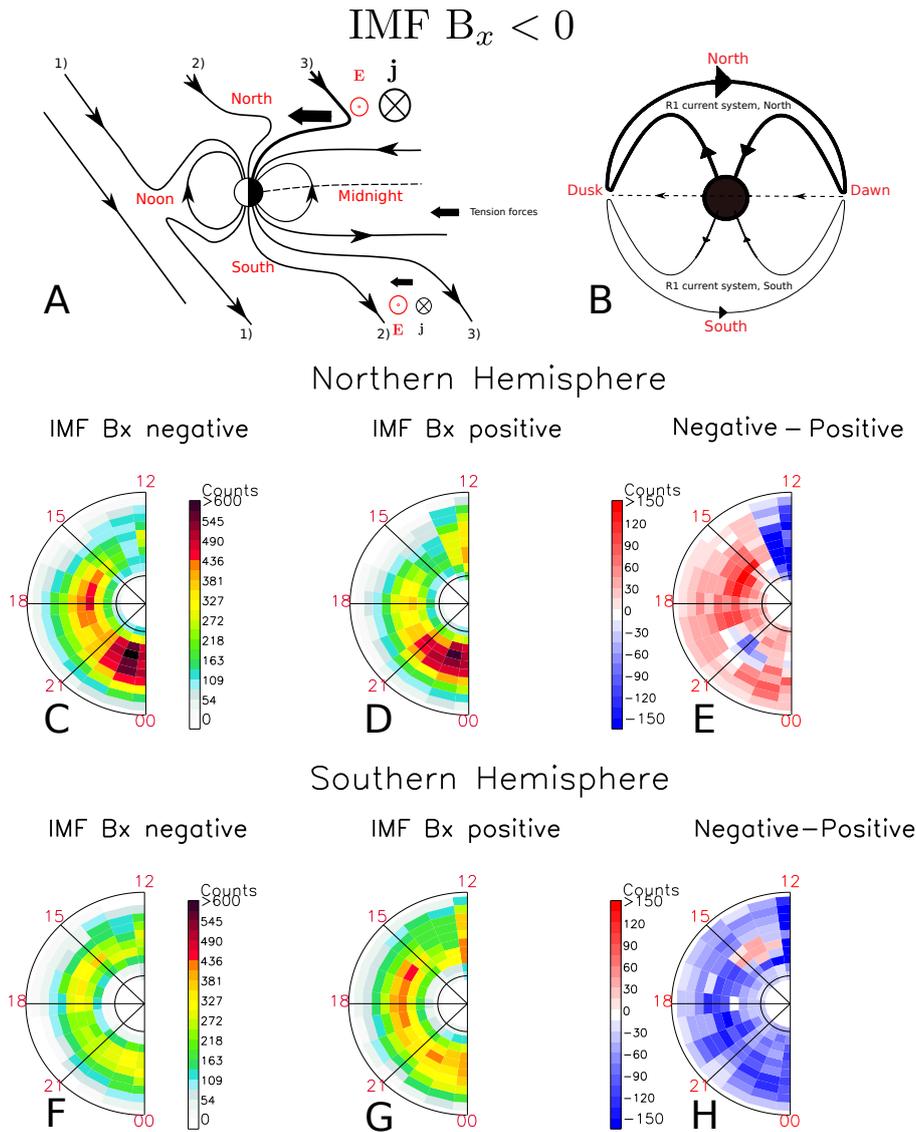

**Figure 11. (a)** Due to a negative IMF $B_x$ (and $B_z < 0$) the magnetic tension force on open field lines (2 and 3) is larger in the Northern Hemisphere (large black arrows) than in the Southern Hemisphere (Cowley, 1981). **(b)** Associated current systems. **(c and f)** Northern and Southern Hemisphere for IMF $B_x$ negative. **(d and g)** Northern and Southern Hemisphere for IMF $B_x$ positive. **(e)** The difference between **(c)** and **(d)**. **(h)** The difference between **(f)** and **(g)**. Panels **(c–e)** are similar to Fig. 2 **(a–c)**, and panels **(f–h)** are similar to Fig. 3 **(a–c)** in Reistad et al. (2014).

In Fig. 12b–d, we illustrate how such a $B_y$ stress in the tail, imposed by the IMF, can propagate to the ionosphere by considering the forces acting on flux tubes. Figure 12b shows the two polar caps (north at the top, south at the bottom) connected by a field line in the mid-tail with foot points in the dawn convection cells. The situation is shown for a positive IMF $B_y$; hence the crescent convection cell is seen on the dawnside in the Northern Hemisphere and on the duskside in the Southern Hemisphere. The asymmetric pressure forces from the lobes, indicated by the $-\nabla P'$ arrows, are in this situation balanced by the tension forces on the field line due to the bending, illustrated by the $\vec{T}$ arrows. The lower part of Fig. 12b shows a view from the side in the $xz$ plane of the same flux tube in the mid-tail. The box in the equatorial plane indicates the region where the field has a $B_y$ component, and for simplicity we use a step function for this $B_y$ field so the currents in the $z$ directions (bottom part Fig. 12b) are only present on the inner and outer edge of this box (purple arrows). The bending of the field due to the asymmetric pressure forces $-\nabla P'$ requires a pair of currents to be present in the $x$ direction within this box. They are indicated by the red and blue arrows in both the upper and lower part of Fig. 12b. In this situation, when the magnetic tension force and the asymmetric magnetic pressure force balance, this configuration will remain stable, the current system will close locally, and stress will not be transported to the ionosphere.

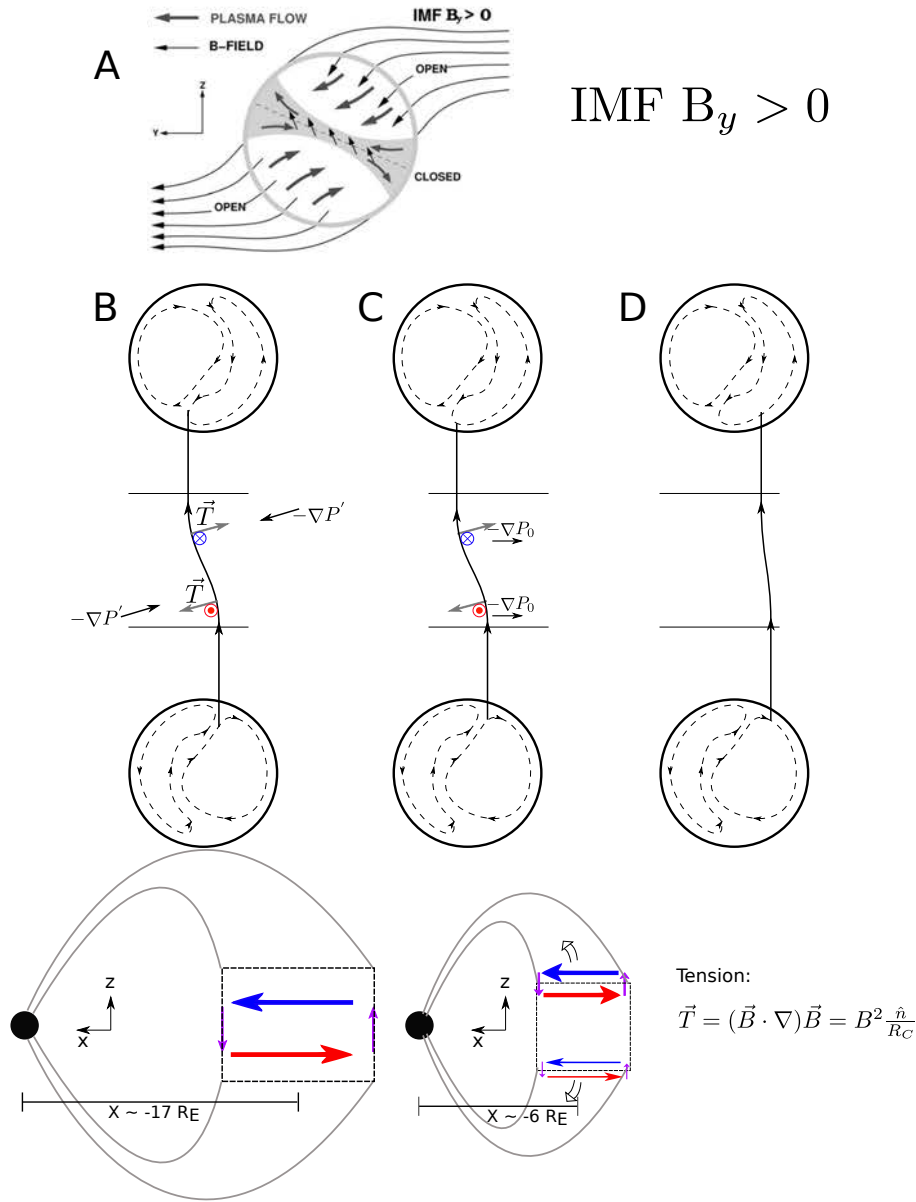

**Figure 12.** (a) Asymmetric entry of magnetic flux in the lobes during positive IMF $B_y$ conditions (Fig. 3a from Liou and Newell, 2010). (b–d) Evolution of a flux tube on closed field lines with asymmetric foot points in the dawn convection cell during IMF $B_y$ positive conditions. Upper panels show pressure, tension and asymmetric foot points into the dawn cells. Lower panels show the associated current systems seen from dusk. (b) In the mid-tail region the asymmetric pressure forces due to IMF $B_y$ and the magnetic tension forces on the flux tube are balanced. Currents close locally as indicated in the lower panel. (c) At a later stage the flux tube moves earthward and is affected by the (total) pressure gradients surrounding the Earth (plasma and magnetic field). Now the forces do not balance. In the Northern Hemisphere these forces point in the same direction. Hence, most of the stress is transmitted into this hemisphere and the northern foot point will catch up with the southern counterpart to restore symmetry, as seen in (d).

As the flux tube convects closer to the Earth the asymmetric lobe pressure will become less significant and the flux tube will rather feel the pressure from the Earth's magnetic field, $-\nabla P_0$. This is illustrated in Fig. 12c. For a flux tube with foot points in the dawn cells, the magnetic pressure force from the Earth's magnetic field ($-\nabla P_0$) will act dawnward (and tailward) along the entire flux tube. This means that the pressure force will act in the same direction as the tension force in the Northern Hemisphere and opposite to the tension force in the Southern Hemisphere. Consequently, most of the stress is transmitted towards the Northern Hemisphere, as illustrated in the lower part of Fig. 12c. This will act to restore symmetry of the foot points on the flux tube. Therefore, the Northern Hemisphere foot point will move faster than the southern hemispheric end to restore symmetry. This is what is seen in Fig. 12d.

As the stress propagates, mostly to the Northern Hemisphere from the situation in panels c to d, it can be represented as a field-aligned current going from the equatorial plane to the northern ionosphere. This propagation is illustrated in the lower part in Fig. 12c. Hence, we cannot call this an interhemispheric current, although the direction of this current (purple arrows) is consistent with what was sketched by Stenbaek-Nielsen and Otto (1997). Furthermore, we would expect to see the signature post-midnight in the Northern Hemisphere. If we had considered a flux tube convecting earthward on the dusk cell and using the same argument, we would expect the stress and the field-aligned current to be transmitted primarily to the Southern Hemisphere. The directions of these currents are also consistent with Stenbaek-Nielsen and Otto (1997).

Two important distinctions can be made from this scenario: (1) IMF $B_y$ does not penetrate the magnetosphere, but through asymmetric lobe pressure it induces a $B_y$ component (with the same sign as IMF $B_y$ in the closed magnetosphere). (2) The currents are not interhemispheric but rather asymmetric from the plasma sheet into the two hemispheres. More explanations, model results and interpretation on how this IMF $B_y$-induced scenario works can be found in Tenfjord et al. (2015).

### 4.3 "Banana" current

The eastward current on the inner edge of the plasma pressure peak is not always evenly distributed in local time, and therefore this current is carried not only by the eastward symmetric ring current but also by some other current system. The eastward current is a magnetization current flowing along plasma pressure isocontours (Roelof, 1989). When the plasma pressure is symmetric in local time, then so is this magnetization current, creating the eastward and westward symmetric ring currents. When the plasma pressure has lo-calized peaks, magnetization currents then flow around each one of these peaks. The portion of the current that flows completely around the localized pressure peak, which accounts for all of the asymmetric eastward current, is called the banana current. Because of the decreasing magnetic field with radial distance, the outer westward current is always larger than the eastward current, and this unbalanced magnetization current closes through the ionosphere as the partial ring current. Figure 13 presents a schematic view on this current system.

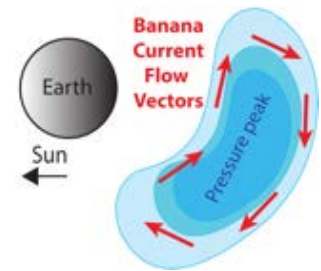

**Figure 13.** Schematic view of banana current system in the equatorial plane.

Liemohn et al. (2013a) conducted a systematic analysis of the asymmetric eastward current, concluding that it closes in a loop around a localized pressure peak, with an outer westward component flowing in the same direction and MLT extent as the partial ring current. They also concluded, from numerical modeling results, that this current intensifies in the main phase just prior to the partial ring current, with a peak magnitude of several mega-amps. By the late recovery phase, the eastward symmetric ring current is typically larger than this current system, although both current systems are quite small by that time (i.e., much less than 1 MA).

This current system was noted by Roelof (1989) and Roelof et al. (2004) in current loop calculations derived from energetic neutral atom (ENA) images. They first inverted the observed ENA emissions into an ion flux distribution in the equatorial plane, and then integrated the result across all energy channels to obtain a pressure map in the inner magnetosphere, revealing a localized pressure peak for their selected interval. From this, they calculated the perpendicular current vectors at each location and subsequently the field-aligned currents. They traced current loops through this vector field and demonstrated that a current system exists that flows around the pressure peak. These current loops were equatorward of the partial ring current, which only flowed in the westward direction around the outside of the pressure peak and then via field-aligned currents to the ionosphere. It was briefly mentioned in several modeling studies (e.g., Liemohn et al., 2011, 2013b). They calculated current traces from MHD results and noted the existence of current loops flowing around nightside localized pressure peaks.

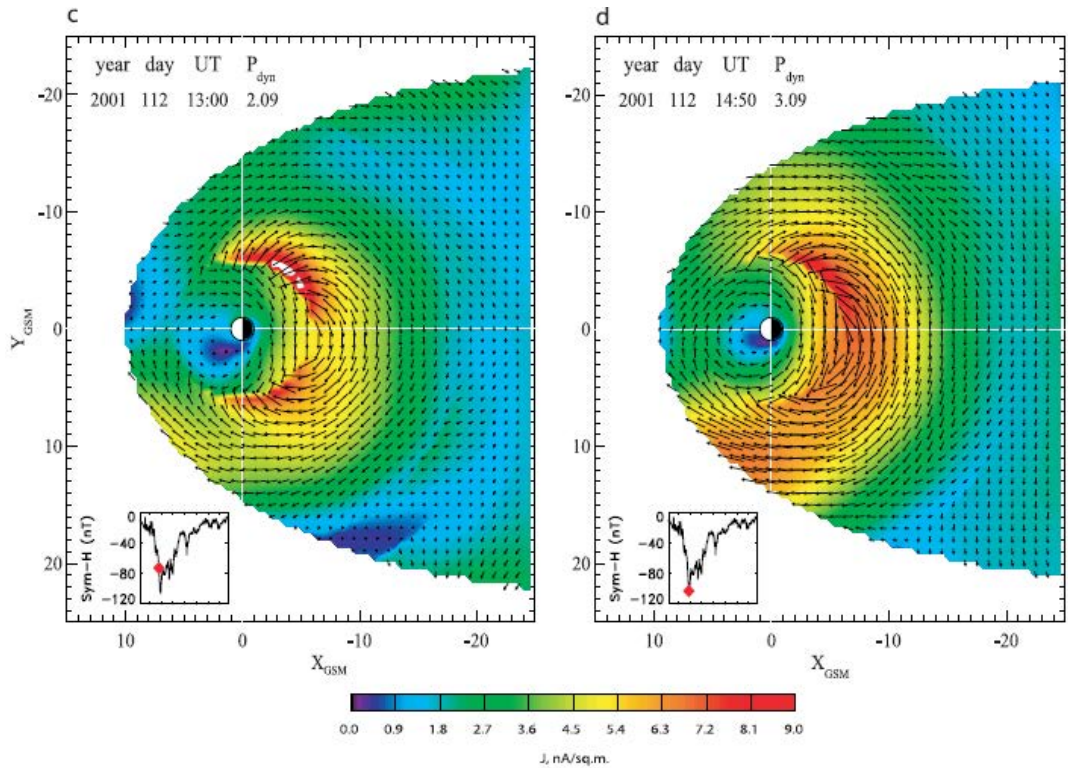

**Figure 14.** Modeled equatorial current density during the April 2001 storm main phase (Fig. 4c and d from Sitnov et al., 2008).

Furthermore, Liemohn et al. (2013a, 2015) systematically investigated the asymmetric eastward current of the inner magnetosphere, including an examination of its expected magnitude and associated magnetic perturbation. They found that this current system reached $\sim 6$ MA during the main phase of an intense magnetic storm. In addition, the timing of this peak is between the maxima of the two other asymmetric current systems: i.e., the progression goes from tail current dominance at the beginning of the main phase to banana current dominance and then to partial ring current dominance in the late main and early recovery phases. The symmetric ring current was the largest near-Earth nightside current system in the late recovery phase.

### 4.4 Duskside tail-like current during storm main phase

The conventional current systems may change their location and intensity and even topology during geomagnetic storms. The advanced empirical modeling by Tsyganenko and Sitnov (2007) and Sitnov et al. (2008) has revealed the strong equatorial westward current on the duskside during the main phase. In contrast to conventional partial ring current, this current concentrates near the neutral sheet region and may have a half-thickness less than 1 $R_E$ (Dubyagin et al., 2013a). In this respect, its geometry is closer to the cross-tail current. On the other hand, it may close through the ionosphere (Dubyagin et al., 2013b) like partial ring current. However, its closure path can change during the course of the main phase (Sitnov et al., 2008). Figure 14 shows the snapshots of the equatorial current density of the Sitnov et al. (2008) (Fig. 4c and d) empirical model for two moments during a moderate storm. The arrows in Fig. 14 show the projection of the current density vectors onto the equatorial plane, and the color shows the magnitude of this projection. The divergence and convergence of the arrows correspond to the current following from the ionosphere to the equator and vice versa, respectively. The red point in the insert shows the time of the current density snapshot with respect to the SYM-H index variation. Figure 14, right panel, shows that the current flows out of the ionosphere as R2 FAC in the post-midnight sector and closes through the dayside magnetopause in the evening sector during the storm peak. On the other hand, Fig. 14, left panel, shows that during the main phase it closes almost entirely through the ionosphere.

Although it was speculated by Sitnov et al. (2008) that this current was associated with ion outflow found in the numerical kinetic simulations (Ebihara, and Ejiri, 1998; Liemohn et al., 1999), it is not immediately obvious, since the electric current stream lines are not directly related to the plasma flow lines. The convective drift in electric field does not produce electric current at all. Although the ions are main carriers of the cross-field current in the inner magnetosphere and their drift does create electric current as well as mass trans-port, this drift current represents only part of the total current, which is a sum of the drift and magnetization currents.

### 4.5 Eastward current at 5–6 $R_E$

Occasionally, multiple injections from the plasma sheet into the inner magnetosphere will occur near enough in time that the pressure peaks from both injections will coexist in near-Earth space but occur far enough apart in time that the pressure peaks do not merge into a single morphological structure. This is clearly seen in the pressure profile of Fig. 1 from Lui et al. (1987) obtained from AMPTE/CCE data, and in Fig. 2 of Liemohn and Jazowski (2008), obtained from numerical modeling, for example. The result is two relative maxima in plasma pressure as a function of radial distance. Each pressure peak will have an associated magnetization current, and therefore an asymmetric eastward ring current (the banana current, as defined above in Sect. 4.3). This means that there will be two regions of eastward current in the inner magnetosphere, one at the innermost edge of the plasma pressure and another farther out, past a region of westward current, often around 5–6 $R_E$ in equatorial plane radial distance.

Theoretically, there is no limit to the number of distinct eastward current regions in the inner magnetosphere, but there is a practical limit. If the injections from the tail are too close in space or time, then they will merge and not create an additional eastward current in the inner magnetosphere. Conversely, if the injections are two far apart in space or time, then they will not radially coexist and therefore not form the double eastward current system. Because the drift speeds of keV-energy ions inside of geosynchronous orbit are on the order of several Earth radii per hour, an optimal injection cadence to observe the extra eastward current is about 1–2 h in UT (e.g., Liemohn et al., 2011).

### 4.6 Cut-ring current

Antonova and Ganushkina (2000) and Antonova (2003, 2004) assume that the high-latitude continuation of the ordinary ring current exists which splits into two branches in the dayside magnetosphere. This current was named the cut-ring current (CRC) and it was suggested that it can be generated by the plasma pressure gradients directed earthward.

Figure 15 shows an example of the calculation of isolines of minimal values of the magnetic field using the TS05 Tsyganenko and Sitnov (2005) model. The magnetic field minima are above and below the equatorial plane near noon. This structure of the magnetic field determines the drift trajectories of particles. Values of current densities and integral transverse current can be estimated assuming the magnetostatic equilibrium when distribution of plasma pressure is nearly isotropic, which was observed beyond geostationary distances (De Michelis et al., 1999). Transverse current can be obtained from Eq. (10), which indicates that the plasma surrounding the Earth contains transverse westward current when the plasma pressure gradient is earthward.

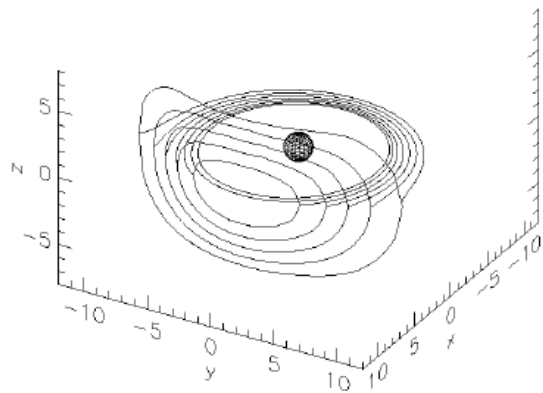

**Figure 15.** The configuration of $B = $ const isolines, calculated using the TS05 Tsyganenko and Sitnov (2005) model with quiet-time parameters (Fig. 1 from Antonova et al., 2009).

The verification of the existence of such a current requires the analysis of global plasma pressure distribution. Antonova et al. (2009) analyzed the radial profiles of plasma pressure gradients obtained from the THEMIS-B satellite data for the period 2 June–29 October 2007 in the equatorial plane near noon at geocentric distances from 7 to 12 $R_E$. The dayside configuration of the geomagnetic field has been obtained using the T02 Tsyganenko (2002) magnetic field model. The estimated value of integral current was $5.8 \times 10^5$ A in both hemispheres. This value was in agreement with estimations obtained by Lui and Hamilton (1992) and De Michelis et al. (1997, 1999).

## 5 Discussion

The definitions in Sects. 3 and 4 provide an excellent resource for understanding the history of the discovery and interpretation of each electric current system in geospace. While these are very useful, in practice, the implementation of defining current systems can be problematic if not handled carefully. Here, we detail a specific example of the many ways to define currents in data and model results, despite the agreement on the basic definitions given above. The specific illustration for consideration here is the near-Earth nightside, a region in which several current systems flow in close proximity, changing in location and intensity throughout geomagnetic activity.

Table 1 presents definitions for three current systems (the symmetric current, the partial ring current, and the tail current) that have all been used in various studies to delineate the current systems within observational or modeling results. The methods are grouped into three categories. The first type of classification includes those methods based on the characteristics of the electric current. For example, they use the closure path, the current intensity or steadiness, or the flow direction of the current. The second classification contains those methods based on the charged particle population. These define the current based on the presence of particles in particular energy ranges or use the characteristics of the energy spectrograms, or properties like the plasma beta value, for defining current regions. The third category includes those methods based solely on spatial location, defining inner and outer edges of the regions for each current system.

Table 1. Definitions for the symmetric ring current, the partial ring current, and the tail current.

| Symmetric ring | Partial ring | Tail |
|---|---|---|
| closed loop around Earth | partly around Earth, then FACs through ionosphere | nightside, then on magnetopause back to dawn |
| all "inner mag" current when it is uniform in MLT | all "inner mag" current when it is not uniform in MLT | all "plasma sheet" current |
| non-fluctuating part of Cluster perigee current | cannot tell apart | fluctuating part of Cluster perigee current |
| SYM-H not from partial RC | ASY-H part of SYM-H | not part of this definition |
| Gaussian peak in $R$ close to Earth | Gaussian in $R$, MLT cosine wave | slabs with peak on $y, z = 0$ line |
| east: all current at $< 4.5\, R_E$, west: $4.5$–$6.5\, R_E$ | cannot tell apart | all current in the $6.5$–$9.5\, R_E$ range |
| inside of geosynch. orbit | cannot tell apart | outside geosynch. orbit |
| all current inside of $8\, R_E$ | cannot tell apart | all current outside $8\, R_E$ |
| all current where $\beta \ll 1$ | cannot tell apart | all current where $\beta > 1$ |
| ion drift bands, $> 10$ keV | cannot tell apart | 1–20 keV plasma sheet |
| ions of $>$ tens of keV | ions of $>$ tens of keV | ions of $< 10$ keV |
| $D_{ENA}$ (dB from P ENA) | cannot tell apart | SYM-H $- D_{ENA}$ |
| anisotropic (trapped) ion populations | cannot tell apart | isotropic ion populations |

What this table highlights is that many different definitions exist for these three well-known current systems. That is, the same phrase, say "symmetric ring current", is used in several different ways depending on the methodology used to define that phrase in that particular study. An author of a report will usually choose only one of these definitions (or perhaps create yet another one not listed). Sometimes this definition is explicitly stated in the report but many times it is not. This assumption that readers will apply the same definition for that phrase leads to confusion. Furthermore, even when the current system definition is clearly stated, those building on the results misinterpret the findings by either not taking into account the limitations of that method or by indiscriminately combining the findings based on different methods.

The advancement of knowledge about the geospace system is really about the physical processes governing its development and evolution in the presence of some initial and/or boundary conditions. This advancement is sometimes best made with the choice of a particular definition for the current systems. Therefore, no particular definitional methodology is advocated over another. Rather, the present paper can serve as a reference for future current definitions in magnetospheric studies, avoiding confusion that could be generated by using ad hoc definitions.

## 6 Conclusions

This paper presents a review of the generally accepted definitions of current systems flowing within geospace. The measurement methods for observing currents were summarized, and the difference between electric currents and charged particle drifts was discussed. Explanations of the dominant current systems were given, followed by descriptions of the transient and unusual current systems within geospace. Finally, the example of the near-Earth nightside region was used to highlight a source of confusion within the field regarding the many possible definitions available for current systems in this region.

The main findings and points to take away are as follows:

1. The measurement of electric currents in space is a difficult process. We hope that this review provides a reference for understanding the various techniques and the applicability and limitations of each.

2. Electric currents are not equivalent to particle guiding-center drifts. In fact, the guiding-center drift motion cancels out of the equation for the total transverse current, with the only contributors being the magnetization current terms of pressure gradients and pressure anisotropies. That is, particles that are all drifting in one direction might actually be creating a current flowing in the opposite direction.

3. Over a dozen different types of current systems have been distinctly identified and named in the magnetosphere–ionosphere system. We hope that this review serves as a reference highlighting the history of discovery and understanding regarding these current systems and as a source for their commonly used definitions.

4. It is crucial for each researcher to carefully and fully define terms related to current systems. Comparisons between studies are greatly hampered when the specific definitions of current systems are assumed and not explicitly declared, while excellent progress has been made when these definitions are presented and incorporated into the interpretation.

*Acknowledgements.* The authors thank the International Space Science Institute in Bern, Switzerland, for their support of an international team on "Resolving Current Systems in Geospace". The work of N. Ganushkina and S. Dubyagin was partly supported by the Academy of Finland. The work of N. Ganushkina was also partly supported by NASA award NNX14AF34G. The part of the research done by N. Ganushkina and S. Dubyagin leading to these results received funding from the European Union Seventh Framework Programme (FP7/2007–2013) under grant agreement 606716 SPACESTORM and from the European Union's Horizon 2020 Research and Innovation programme under grant agreement 637302 PROGRESS. The work at Birkeland Centre for Space Science, University of Bergen, Norway, was supported by the Research Council of Norway/CoE under contract 223252/F50. S. E. Milan was supported by the Science and Technology Facilities Council (UK), grant no. ST/K001000/1. Work at JHU/APL was supported by NSF grant AGS-1104338.

The topical editor C.-P. Escoubet thanks A. Milillo and two anonymous referees for help in evaluating this paper.